Featured in Physics

# GW190412: Observation of a binary-black-hole coalescence with asymmetric masses

R. Abbott *et al.*[*]
(LIGO Scientific Collaboration and Virgo Collaboration)

We report the observation of gravitational waves from a binary-black-hole coalescence during the first two weeks of LIGO's and Virgo's third observing run. The signal was recorded on April 12, 2019 at 05:30:44 UTC with a network signal-to-noise ratio of 19. The binary is different from observations during the first two observing runs most notably due to its asymmetric masses: a $\sim 30\ M_\odot$ black hole merged with a $\sim 8\ M_\odot$ black hole companion. The more massive black hole rotated with a dimensionless spin magnitude between 0.22 and 0.60 (90% probability). Asymmetric systems are predicted to emit gravitational waves with stronger contributions from higher multipoles, and indeed we find strong evidence for gravitational radiation beyond the leading quadrupolar order in the observed signal. A suite of tests performed on GW190412 indicates consistency with Einstein's general theory of relativity. While the mass ratio of this system differs from all previous detections, we show that it is consistent with the population model of stellar binary black holes inferred from the first two observing runs.

## I. INTRODUCTION

The first detections [1–12] of gravitational-wave (GW) signals by the Advanced Laser Interferometer Gravitational-wave Observatory (LIGO) [13] and Advanced Virgo [14] detectors during their first two observing runs have begun to constrain the population of astrophysical binary black holes (BBHs) [15]. Prior to the start of the third observing run (O3) the Advanced LIGO and Advanced Virgo detectors were upgraded to increase the sensitivity of all three interferometers [16–19]. This increase in sensitivity has broadened the detector network's access to GW signals from the population of merging BBH sources [20], allowing for the detection of rarer systems.

To be able to characterize the full range of potential systems, models of the gravitational radiation emitted by BBHs are continuously being improved. In particular, physical effects associated with unequal masses and misaligned spins have recently been extended in models covering the inspiral, merger and ringdown of BBHs [21–33]. For GW signals with sufficient signal-to-noise ratio (SNR), the inclusion of these effects is important to accurately infer the source parameters. In addition, improved signal models allow for stronger tests of the validity of general relativity (GR) as the correct underlying theory.

[*]Full author list given at the end of the article.

In this paper we report the detection of GWs from a BBH whose properties make it distinct from all other BBHs reported previously from the first two observing runs. The event, called GW190412, was observed on April 12, 2019 at 05:30:44 UTC by the Advanced Virgo detector and both Advanced LIGO detectors. While the inferred individual masses of the coalescing black holes (BHs) are each within the range of masses that have been observed before [7,9–12], previously detected binaries all had mass ratios $q = m_2/m_1$ (with $m_1 \geq m_2$) that were consistent with unity [34]. GW190412, however, is the first observation of GWs from a coalescing binary with unequivocally unequal masses, $q = 0.28^{+0.12}_{-0.07}$ (median and 90% symmetric credible interval). The mass asymmetry of the system provides a second novelty of GW190412: the GWs carry subtle, but for the first time clearly measurable, imprints of radiation that oscillates not only at the binary's dominant GW emission frequency, but also at other frequencies with subdominant contributions. We introduce the nature of these corrections and present the source parameters inferred for GW190412 using signal models that include higher multipoles.

This paper is organized as follows: in Sec. II we report details on the detection of GW190412. The source properties are discussed in Sec. III. Section IV presents a suite of analyses illustrating that the observed data indeed contain measurable imprints of higher multipoles. In Sec. V we present tests of GR performed in this previously unexplored region of the parameter space. Implications for our understanding of the BBH population and formation channels are discussed in Secs. VI and VII.

## II. DETECTORS AND DETECTION

The third observing run of LIGO [35] and Virgo [14] began on April 1, 2019, and GW190412 occurred in the second week of the run. At the time of the event, both LIGO detectors and the Virgo detector were online and operating stably for over 3.5 hours. Strain data from around the time of GW190412 for all three detectors is shown in Fig. 1, with excess power consistent with the observed signal present in all detectors. The relative sensitivity of the LIGO and Virgo detectors accounts for the difference in strength of the signal in the data.

LIGO and Virgo interferometers are calibrated using electrostatic fields and radiation pressure from auxiliary lasers at a known frequency and amplitude [37–39]. At the time of the event, the maximum calibration error at both LIGO sites is 7.0% in amplitude and 3.8° in phase. At Virgo, the errors are 5.0% in amplitude and 7.5° in phase.

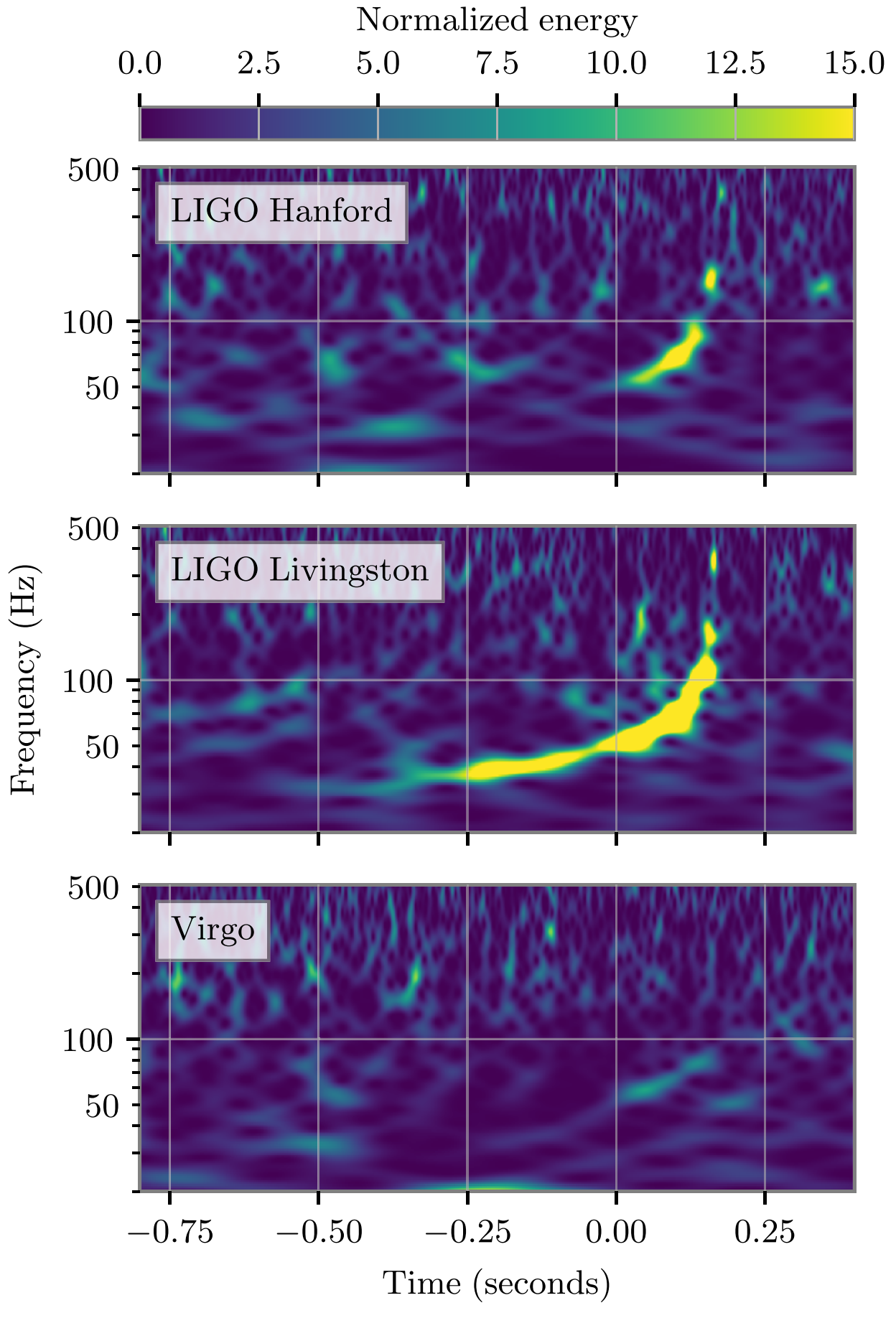

FIG. 1. Time-frequency representations [36] of the strain data at the time of GW190412 in LIGO Hanford (top), LIGO Livingston (middle), and Virgo (bottom). Times are shown from April 12, 2019, 05:30:44 UTC. Excess power, consistent with the measured parameters of the event, is present in all three detectors. The amplitude scale of each time-frequency tile is normalized by the respective detector's noise amplitude spectral density. The lower frequency limit of 20 Hz is the same as in analyses of the source properties of GW190412.

Numerous noise sources that limit detector sensitivity are measured and subtracted, including noise from calibration lines and noise from the harmonics of the power mains. Similar to procedures from the second observing run [40,41], these noise sources are linearly subtracted from the data using auxiliary witness sensors. In O3, this procedure was completed as a part of the calibration pipeline [38], both in low latency and offline. Additional noise contributions due to nonlinear coupling of the 60 Hz power mains are subtracted for offline analyses using coupling functions that rely on machine learning techniques [42].

GW190412 was initially detected by the GstLAL [43], SPIIR [44], CWB [45], MBTA [46], and PyCBC Live [47] pipelines running in low-latency configuration, and reported under the identifier S190412m. The GstLAL, SPIIR, MBTA, and PyCBC Live pipelines identify GW signals by matched filtering [48–50] data using a bank of filter waveforms that cover a wide range of source parameters [51–57]. The coherent, unmodeled CWB pipeline [45], identifies clusters of coherent excess power with increasing frequency over time in data processed with the wavelet transform [58].

All analysis pipelines running in low-latency identified GW190412 as a confident event. The observed SNR from the GstLAL pipeline was 8.6 in LIGO Hanford, 15.6 in LIGO Livingston, and 3.7 in Virgo. GW190412 was identified with consistent SNR across all low-latency pipelines.

An alert [59] announcing the event was publicly distributed 60 minutes after GW190412 through NASA's Gamma-ray Coordinates Network and included an initial sky localization computed using a rapid Bayesian algorithm, BAYESTAR [60], applied to data from all available detectors. This sky localization constrained the position of the event to a 90% credible area of 156 $\mathrm{deg}^2$.

Additional offline analysis of the data from April 8 to April 18 was completed by the GstLAL [61,62] and PyCBC [63,64] matched filtering pipelines, and the coherent, template independent CWB pipeline [45]. The offline analyses utilize an updated version of the calibration of the LIGO data [38] and additional data quality vetoes [65]. The GstLAL pipeline incorporates data from all three detectors, while the PyCBC and CWB pipelines only use data from the two LIGO detectors.

All three offline pipelines identified GW190412 as a highly significant GW event. This event was assigned a false-alarm rate (FAR) of $< 1$ per $1 \times 10^5$ years by the GstLAL pipeline and $< 1$ per $3 \times 10^4$ years by the PyCBC pipeline. The template independent CWB pipeline assigned this event a FAR of $< 1$ per $1 \times 10^3$ years. As GW190412 was identified as more significant than any event in the computed background, the FARs assigned by all offline pipelines are upper bounds.

Validation procedures similar to those used to evaluate previous events [7,66] were used in the case of GW190412 to verify that instrumental artifacts do not affect the analysis of the observed event. These procedures rely upon sensor arrays at LIGO and Virgo to measure environmental disturbances that could potentially couple into the interferometers [67]. For all three interferometers, these procedures identified no evidence of excess power from terrestrial sources that could impact detection or analysis of GW190412. Data from Virgo contains instrumental artifacts from scattered light [68] that impact data below 20 Hz within 4 seconds of the coalescence time. As analyses of GW190412 source properties only use data from above 20 Hz, no mitigation of these artifacts was required.

## III. SOURCE PROPERTIES

### A. On radiative multipoles and source properties

GW radiation is observed as a combination of two polarizations, $h_+$ and $h_\times$ weighted with the detector response functions [69]. For GW theory, it is efficient to work with the complex valued quantity, $h = h_+ - i h_\times$. From the perspective of the observer, $h$ can be expanded into multipole moments using spherical polar coordinates defined in a source centered frame [70]. Each multipole moment encodes information about the gravitationally radiating source. Interfacing GW theory with data analysis allows the connections between radiative multipole moments and source properties to be decoded.

Starting with the pioneering work of Einstein, and later refined by Newman, Penrose, Thorne and many others, GWs have been known to be at least quadrupolar [70–72]. The $-2$ spin-weighted spherical harmonics $_{-2}Y_{\ell m}(\theta,\phi)$ have been found to be the simplest appropriate harmonic basis. They are the orthonormal angular eigenfunctions of gravitationally perturbed spherically symmetric spacetimes and we refer to their beyond-quadrupolar multipole moments in this basis as higher multipoles [72–77].

In this basis the multipolar decomposition is

$$h_+ - i h_\times = \sum_{\ell\geq 2} \sum_{-\ell\leq m\leq \ell} \frac{h_{\ell m}(t,\boldsymbol{\lambda})}{D_\mathrm{L}} \, {}_{-2}Y_{\ell m}(\theta,\phi), \quad (1)$$

where $(\theta,\phi)$ are respectively the polar and azimuthal angles defining the direction of propagation from the source to the observer, and $D_\mathrm{L}$ the luminosity distance from the observer. The radiative multipoles, $h_{\ell m}$, depend on source properties (condensed in $\boldsymbol{\lambda}$) such as the BH masses, $m_1$ and $m_2$, and their spins $\vec{S}_1$ and $\vec{S}_2$.

When at least one of the BH spins is misaligned with respect to the binary's angular momentum, the orbital plane and the spins precess around the direction of the total angular momentum. We refer to these systems as precessing. For nonprecessing systems, reflection symmetry about a fixed orbital plane results in a complex conjugate symmetry between moments of azimuthal index $m$ and $-m$. Therefore, when we refer to a specific $(\ell, m)$ multipole, $h_{\ell m}$, we always mean $(\ell, \pm m)$. For precessing systems and their nonfixed orbital plane, one may define a coprecessing frame such that $\theta$ is relative to the direction of maximal radiation emission. In this coprecessing frame $h_{\ell m}$ approximately take on features of their nonprecessing counterparts [78–82].

In the early inspiral, the instantaneous frequencies, $f_{\ell m}$, of each $h_{\ell m}$ are linked to the orbital frequency of the binary $f_\mathrm{orb}$ by $f_{\ell m} \simeq m f_\mathrm{orb}$ [74]. In the moments shortly before the two BHs merge, strong-field effects cause this simple scaling to be broken, but its imprint persists through the final stages of coalescence [22]. Therefore, higher multipoles imply an approximately harmonic progression of frequencies within GW signals from quasicircularly coalescing BHs.

The source geometry depends on mass ratio and is most prominently manifested in the relative contribution of multipoles with odd or even $m$: for an exactly equal mass binary with nonspinning components, only multipoles with even $m$ respect orbital symmetry and so are present in the radiation [74]. In this and nearly equal mass cases, the quadrupole, $h_{22}$, is by far the most dominant, followed by other multipoles with even $m$. However, for sufficiently unequal mass ratios, numerical and analytical studies have shown that the $\ell = m = 3$ and subsequent multipoles with $\ell = m$ gain increasing importance [74,75,83–87]. In the context of source detection and characterization, analytical and numerical studies [76,88–97] have shown that higher multipoles can have increasing relative importance as system asymmetry increases due to source properties such as unequal masses, spin magnitudes and, in the case of precession, spin directions.

Source orientation also contributes to higher multipole content and can impact the inference of source parameters. Systems whose orbital planes point directly towards the observer present signals which are generally dominated by $h_{22}$. In such instances the net dependence on $\theta$ and $D_\mathrm{L}$ enters collinearly as $_{-2}Y_{22}(\theta,\phi)/D_\mathrm{L}$. Thus distance and inclination can be approximately degenerate for each GW detector. However, for inclined systems, the higher multipoles introduce other dependencies on $\theta$ via their harmonics, $_{-2}Y_{\ell m}(\theta,\phi)$. Consequently, higher multipoles may break the inclination-distance degeneracy, thereby tightening constraints on inferred source inclination and luminosity distance. We show in Sec. III D how signal models including higher multipoles can break this degeneracy and improve our ability to infer the source parameters of GW190412. When we refer to the inclination of GW190412 below, we define it as the angle between the system's total angular momentum and the direction of propagation from the source to the observer, and denote it by $\theta_{JN}$ for clarity.

Importantly, GW signal models are used to map between radiative multipole content and source properties, and for the robust estimation of source properties, experimental uncertainty requires GW signal models to be used in conjunction with methods of statistical inference.

### B. Method and signal models

We perform an inference of the properties of the binary using a coherent Bayesian analysis of 8 seconds of data around the time of the detection from the three detectors (see e.g., [98]). Results presented here were mainly produced with a highly parallelized version of the code package Bilby [99–101]. Additional analyses were performed with the package LALInference [102]. RIFT [103,104] was also used to check consistency of the intrinsic parameters and for corroborating the Bayes factors that are presented below. The power spectral density (PSD) of the noise that enters the likelihood calculation is estimated from the data using BayesWave [105,106]. The low-frequency cutoff for the likelihood integration is set to 20 Hz, and the prior distributions we use are described in Appendix B 1 of [7]. We note that after initial analyses with large prior intervals on the individual masses, more restrictive prior boundaries were introduced to accelerate further Bilby analyses. Those additional boundaries do not affect the posterior probabilities reported below as the likelihood is insignificantly small outside the allowed prior region. However, Kullback-Leibler divergences [107] calculated between prior and posterior are sensitive to this choice. Full prior specifications can be found in the data accompanying this paper [108].

TABLE I. Waveform models used in this paper. We indicate which multipoles are included for each model. For precessing models, the multipoles correspond to those in the coprecessing frame.

| Family | Short name | Full name | Precession | Multipoles $(\ell, \|m\|)$ | Reference |
|---|---|---|---|---|---|
| EOBNR | EOBNR | SEOBNRv4_ROM | ✗ | (2, 2) | [57] |
| | EOBNR HM | SEOBNRv4HM_ROM | ✗ | (2, 2), (2,1), (3, 3), (4, 4), (5, 5) | [26,32] |
| | EOBNR P | SEOBNRv4P | ✓ | (2, 2), (2, 1) | [33,118,119] |
| | EOBNR PHM | SEOBNRv4PHM | ✓ | (2, 2), (2, 1), (3, 3), (4, 4), (5, 5) | [33,118,119] |
| Phenom | Phenom | IMRPhenomD | ✗ | (2, 2) | [120,121] |
| | Phenom HM | IMRPhenomHM | ✗ | (2, 2), (2, 1), (3, 3), (3, 2), (4, 4), (4, 3) | [22] |
| | Phenom P | IMRPhenomPv2/v3[a] | ✓ | (2, 2) | [23,122] |
| | Phenom PHM | IMRPhenomPv3HM | ✓ | (2, 2), (2, 1), (3, 3), (3, 2), (4, 4), (4, 3) | [24] |
| NR surrogate | NRSur HM | NRHybSur3dq8 | ✗ | $\ell \leq 4$, (5, 5) but not (4, 0), (4, 1) | [27] |

[a]The recently improved, precessing model IMRPhenomPv3 is used in Sec. IVA to calculate Bayes factors. For consistency with previous analyses and computational reasons, the tests presented in Sec. VA use IMRPhenomPv2 instead.

The signal models we use to sample the BBH parameter space are enhanced versions of the models that have been used in past analyses (e.g., [7]). We employ models from the effective-one-body (EOB) [109–112] family that are constructed by completing an analytical inspiral-merger-ringdown description which builds on post-Newtonian (PN) [113–115] and black-hole perturbation theory, with numerical-relativity information. The phenomenological family [116,117] on the other hand, is based on a frequency-domain description of hybridized EOB-inspiral and numerical-relativity merger. The latest developments used here include the effects of higher multipoles in precessing models both in the EOBNR family (SEOBNRv4PHM, [33,118,119]) and the phenomenological family (IMRPhenomPv3HM, [23,24]).

All model variants that we use in the analysis of GW190412 are detailed in Table I. In order to test for imprints of spin precession and higher multipoles in the data, we also perform analyses using models without spin precession and/or without higher multipoles. To verify the robustness of our results against waveform systematics we also performed an analysis using the numerical-relativity surrogate NRHybSur3dq8 [27] that includes the effect of higher multipoles, but is limited to spins aligned with the angular momentum. This surrogate model is constructed from numerical-relativity waveforms extended with EOB-calibrated PN waveforms.

### C. Masses

In Table II we summarize the inferred values of the source parameters of GW190412. The statistical uncertainty is quantified by the equal-tailed 90% credible intervals about the median of the marginalized one-dimensional posterior distribution of each parameter. We report the results obtained with the two most complete signal models—those members of the EOBNR and Phenom family that include both the effects of precession and higher multipoles (see Table I). As a conservative estimate, and because we do not favor one model over the other, we combine the posteriors of each model with equal weight, which is equivalent to marginalizing over a discrete set of signal models with a uniform probability. The resulting values are provided in the last column of Table II, and we refer to those values in the text unless explicitly stated otherwise.

The component masses for this system in the source frame are $m_1 = 30.1^{+4.6}_{-5.3}\ M_\odot$ and $m_2 = 8.3^{+1.6}_{-0.9}\ M_\odot$. They are consistent with the BH mass ranges of population models inferred from the first two LIGO and Virgo observing runs [7]. However, GW190412 is particularly interesting for its measured mass ratio of $q = 0.28^{+0.12}_{-0.07}$. Figures 2 and 3 illustrate that the mass ratio inferred for GW190412 strongly disfavors a system with comparable masses. We exclude $q > 0.5$ with 99% probability. In Sec. VI we show that the asymmetric component mass measurement is robust when analyzed using a prior informed by the already-observed BBH population.

The posteriors shown in Fig. 2 for the two precessing, higher multipole models are largely overlapping, but differences are visible. The EOBNR PHM model provides tighter constraints than Phenom PHM, and the peak of the posterior distributions are offset along a line of high correlation in the $q$–$\chi_\mathrm{eff}$ plane. This mass-ratio–spin degeneracy arises because inspiral GW signals can partly compensate the effect of a more asymmetric mass ratio with a higher effective spin [128–132]. The effective spin [111,133,134] is the mass-weighted sum of the individual spin components $\vec{S}_1$ and $\vec{S}_2$ perpendicular to the orbital plane, or equivalently projected along the direction of the Newtonian orbital angular momentum, $\vec{L}_N$,

$$\chi_\mathrm{eff} = \frac{1}{M}\left(\frac{\vec{S}_1}{m_1} + \frac{\vec{S}_2}{m_2}\right) \cdot \frac{\vec{L}_N}{\|\vec{L}_N\|}, \quad (2)$$

with $M = m_1 + m_2$.

TABLE II. Inferred parameter values for GW190412 and their 90% credible intervals, obtained using precessing models including higher multipoles.

| Parameter[a] | EOBNR PHM | Phenom PHM | Combined |
|---|---|---|---|
| $m_1/M_\odot$ | $31.7^{+3.6}_{-3.5}$ | $28.1^{+4.8}_{-4.3}$ | $30.1^{+4.6}_{-5.3}$ |
| $m_2/M_\odot$ | $8.0^{+0.9}_{-0.7}$ | $8.8^{+1.5}_{-1.1}$ | $8.3^{+1.6}_{-0.9}$ |
| $M/M_\odot$ | $39.7^{+3.0}_{-2.8}$ | $36.9^{+3.7}_{-2.9}$ | $38.4^{+3.8}_{-3.9}$ |
| $\mathcal{M}/M_\odot$ | $13.3^{+0.3}_{-0.3}$ | $13.2^{+0.5}_{-0.3}$ | $13.3^{+0.4}_{-0.4}$ |
| $q$ | $0.25^{+0.06}_{-0.04}$ | $0.31^{+0.12}_{-0.07}$ | $0.28^{+0.12}_{-0.07}$ |
| $M_\mathrm{f}/M_\odot$ | $38.6^{+3.1}_{-2.8}$ | $35.7^{+3.8}_{-3.0}$ | $37.3^{+3.8}_{-4.0}$ |
| $\chi_\mathrm{f}$ | $0.68^{+0.04}_{-0.04}$ | $0.67^{+0.07}_{-0.07}$ | $0.67^{+0.06}_{-0.05}$ |
| $m_1^\mathrm{det}/M_\odot$ | $36.5^{+4.2}_{-4.2}$ | $32.3^{+5.7}_{-5.2}$ | $34.6^{+5.4}_{-6.4}$ |
| $m_2^\mathrm{det}/M_\odot$ | $9.2^{+0.9}_{-0.7}$ | $10.1^{+1.6}_{-1.2}$ | $9.6^{+1.7}_{-1.0}$ |
| $M^\mathrm{det}/M_\odot$ | $45.7^{+3.5}_{-3.3}$ | $42.5^{+4.4}_{-3.7}$ | $44.2^{+4.4}_{-4.7}$ |
| $\mathcal{M}^\mathrm{det}/M_\odot$ | $15.3^{+0.1}_{-0.2}$ | $15.2^{+0.3}_{-0.2}$ | $15.2^{+0.3}_{-0.1}$ |
| $\chi_\mathrm{eff}$ | $0.28^{+0.06}_{-0.08}$ | $0.22^{+0.08}_{-0.11}$ | $0.25^{+0.08}_{-0.11}$ |
| $\chi_\mathrm{p}$ | $0.31^{+0.14}_{-0.15}$ | $0.31^{+0.24}_{-0.17}$ | $0.31^{+0.19}_{-0.16}$ |
| $\chi_1$ | $0.46^{+0.12}_{-0.15}$ | $0.41^{+0.22}_{-0.24}$ | $0.44^{+0.16}_{-0.22}$ |
| $D_\mathrm{L}/\mathrm{Mpc}$ | $740^{+120}_{-130}$ | $740^{+150}_{-190}$ | $740^{+130}_{-160}$ |
| $z$ | $0.15^{+0.02}_{-0.02}$ | $0.15^{+0.03}_{-0.04}$ | $0.15^{+0.03}_{-0.03}$ |
| $\hat{\theta}_{JN}$ | $0.71^{+0.23}_{-0.21}$ | $0.71^{+0.39}_{-0.27}$ | $0.71^{+0.31}_{-0.24}$ |
| $\rho_\mathrm{H}$ | $9.5^{+0.1}_{-0.2}$ | $9.5^{+0.2}_{-0.3}$ | $9.5^{+0.1}_{-0.3}$ |
| $\rho_\mathrm{L}$ | $16.2^{+0.1}_{-0.2}$ | $16.1^{+0.2}_{-0.3}$ | $16.2^{+0.1}_{-0.3}$ |
| $\rho_\mathrm{V}$ | $3.7^{+0.2}_{-0.5}$ | $3.6^{+0.3}_{-1.0}$ | $3.6^{+0.3}_{-0.7}$ |
| $\rho_\mathrm{HLV}$ | $19.1^{+0.2}_{-0.2}$ | $19.0^{+0.2}_{-0.3}$ | $19.1^{+0.1}_{-0.3}$ |

[a]Symbols: $m_i$: individual mass; $M = m_1 + m_2$; $\mathcal{M} = (m_1 m_2)^{3/5} M^{-1/5}$; superscript "det" refers to the detector-frame (redshifted) mass, while without superscript, masses are source-frame masses, assuming a standard cosmology [123] detailed in Appendix B of [7]; $q = m_2/m_1$; $M_\mathrm{f}, \chi_\mathrm{f}$: mass and dimensionless spin magnitude of the remnant BH, obtained through numerical-relativity fits [124–127]; $\chi_\mathrm{eff}$, $\chi_\mathrm{p}$: effective and precessing spin parameter; $\chi_1$: dimensionless spin magnitude of more massive BH; $D_\mathrm{L}$: luminosity distance; $z$: redshift; $\hat{\theta}_{JN}$: inclination angle (folded to $[0, \pi/2]$); $\rho_X$ matched-filter SNRs for the Hanford, Livingston and Virgo detectors, indicated by subscript. $\rho_\mathrm{HLV}$: network SNR.

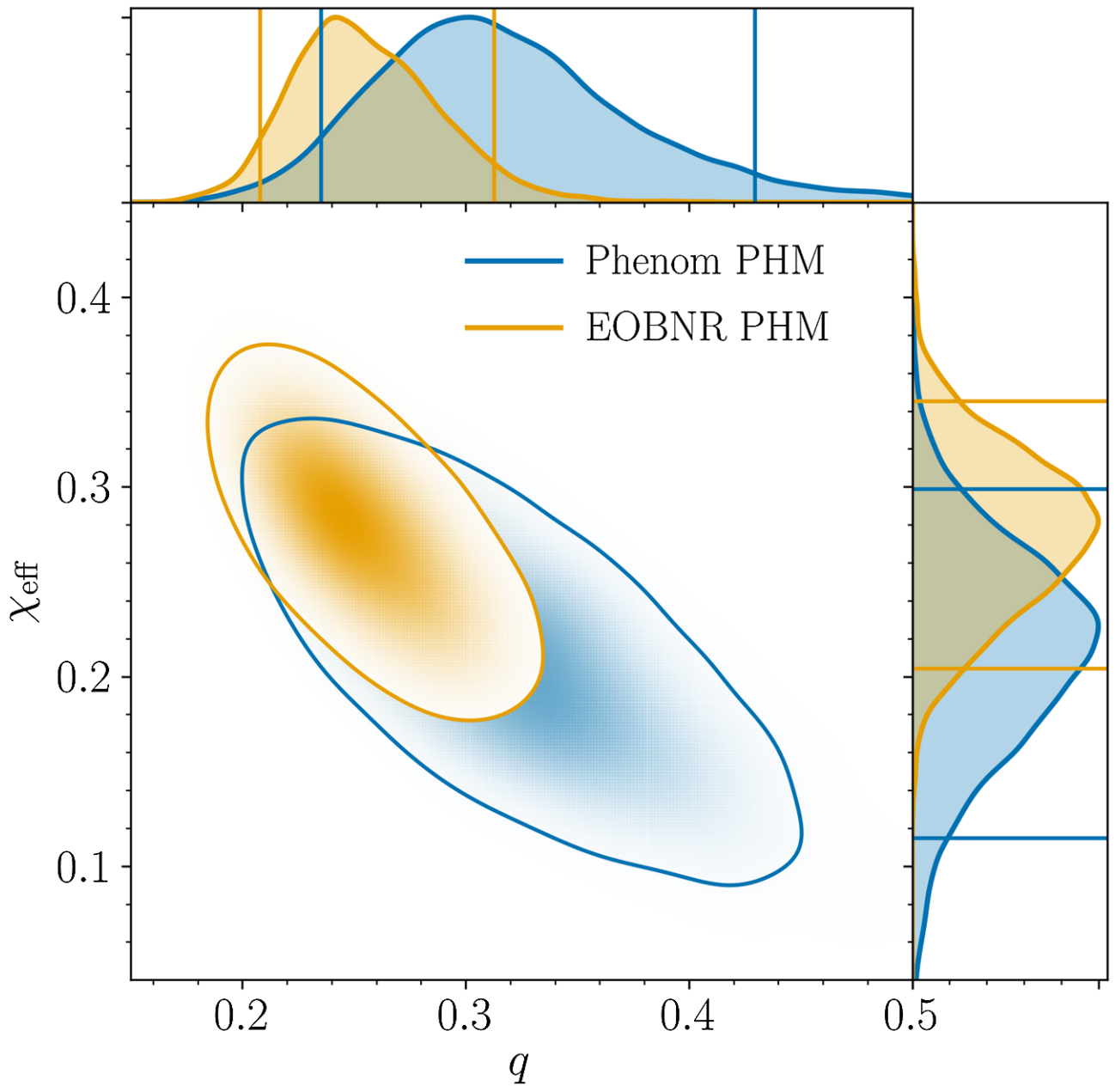

FIG. 2. The posterior distribution for the mass ratio $q$ and effective spin $\chi_\mathrm{eff}$ of GW190412. We show the two-dimensional marginalized distribution as well as the one-dimensional marginalized distribution of each parameter along the axes for two different signal models that each include the effects of precessing spins and higher multipoles. The indicated two-dimensional area as well as the horizontal and vertical lines along the axes, respectively, indicate the 90% credible regions.

GW190412 is in a region of the parameter space that has not been accessed through observations before; and we find that the two models give slightly different, yet largely consistent results. However, this is the first time that systematic model differences are not much smaller than statistical uncertainties. We tested the origin of these differences by repeating the analysis with an extended suite of signal models, as shown in Fig. 3. The results indicate that the mass-ratio measurement of GW190412 is robust against modeling systematics, and the different treatments of higher multipoles in the EOBNR and Phenom families may account for some of the observed differences. We also see that the NRSur HM model and the EOBNR HM model agree well with each other, while the Phenom HM model deviates slightly. This is consistent with the fact that the NRSur HM and EOBNR HM models have some features in common. In NRSur HM, the PN inspiral part of the waveform is calibrated to EOB waveforms, and in EOBNR HM the merger and ringdown part of the waveform is calibrated to a subset of the numerical-relativity simulations used in the construction of NRSur HM. Phenom HM, on the other hand, is based upon an uncalibrated EOB inspiral phase, and higher multipoles are added without any additional numerical-relativity tuning. Further studies will be needed to fully understand the systematics visible here and mitigate them as models improve.

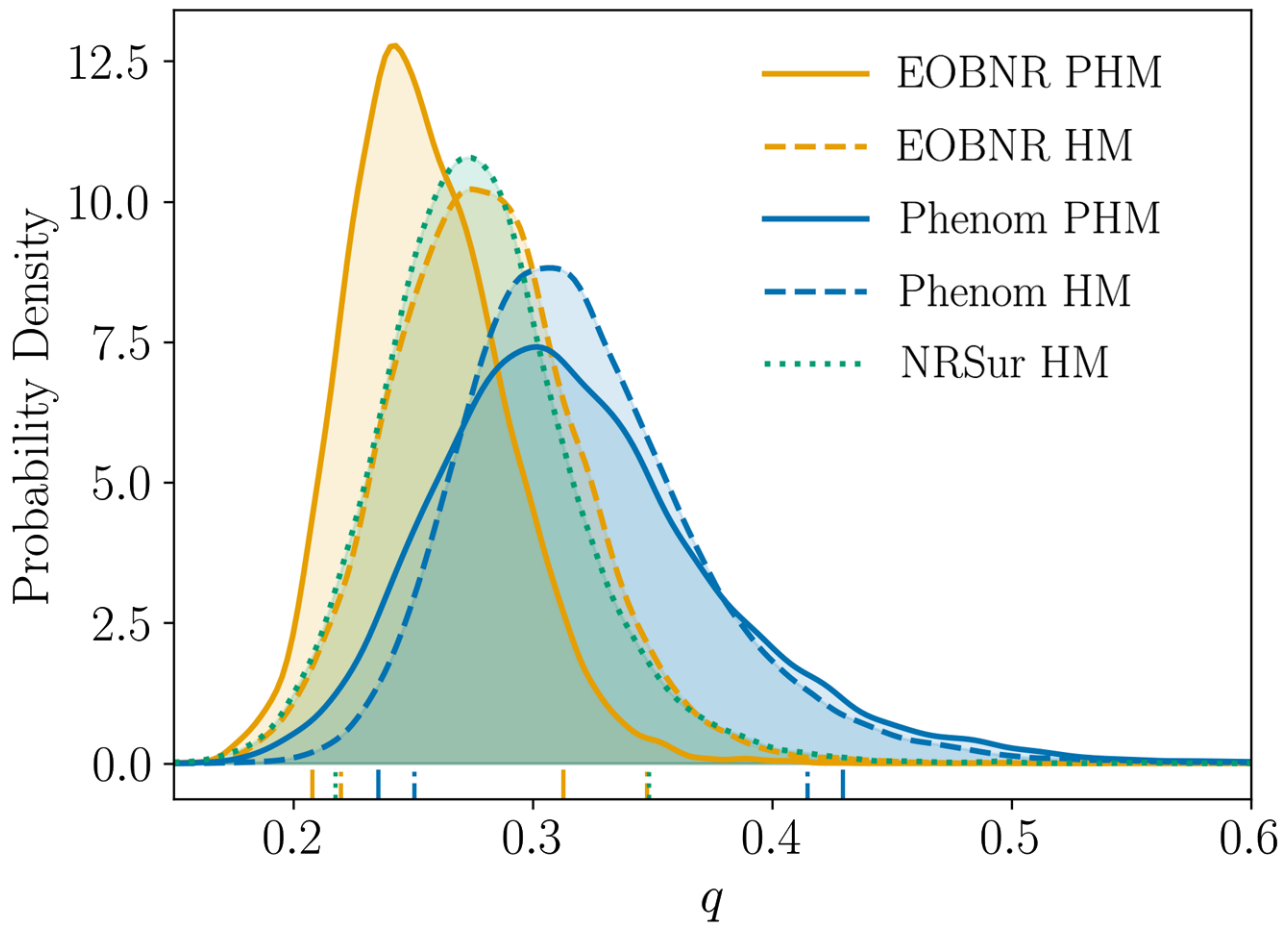

FIG. 3. The one-dimensional posterior probability density for the mass ratio $q$ of GW190412, obtained with a suite of different signal models. The vertical lines above the bottom axes indicate the 90% credible bounds for each signal model.

### D. Orientation and spins

The contribution of higher multipoles in the gravitational waveform is important for the parameter estimation of systems with small mass ratios [135,136]. In Fig. 4 we show the marginalized two-dimensional posterior distribution for luminosity distance and inclination obtained using signal models either without higher multipoles, with higher multipoles, or with higher multipoles and spin precession. The degeneracy between luminosity distance and inclination angle that is present in the results obtained without higher multipoles is broken when higher multipoles are included. The inclusion of precession effects helps to constrain the 90% credible region further. Results obtained with the Phenom family show the same degeneracy breaking when higher multipoles are included, but the 90% credible region obtained with Phenom PHM has some remaining small support for $\theta_{JN} > \pi/2$.

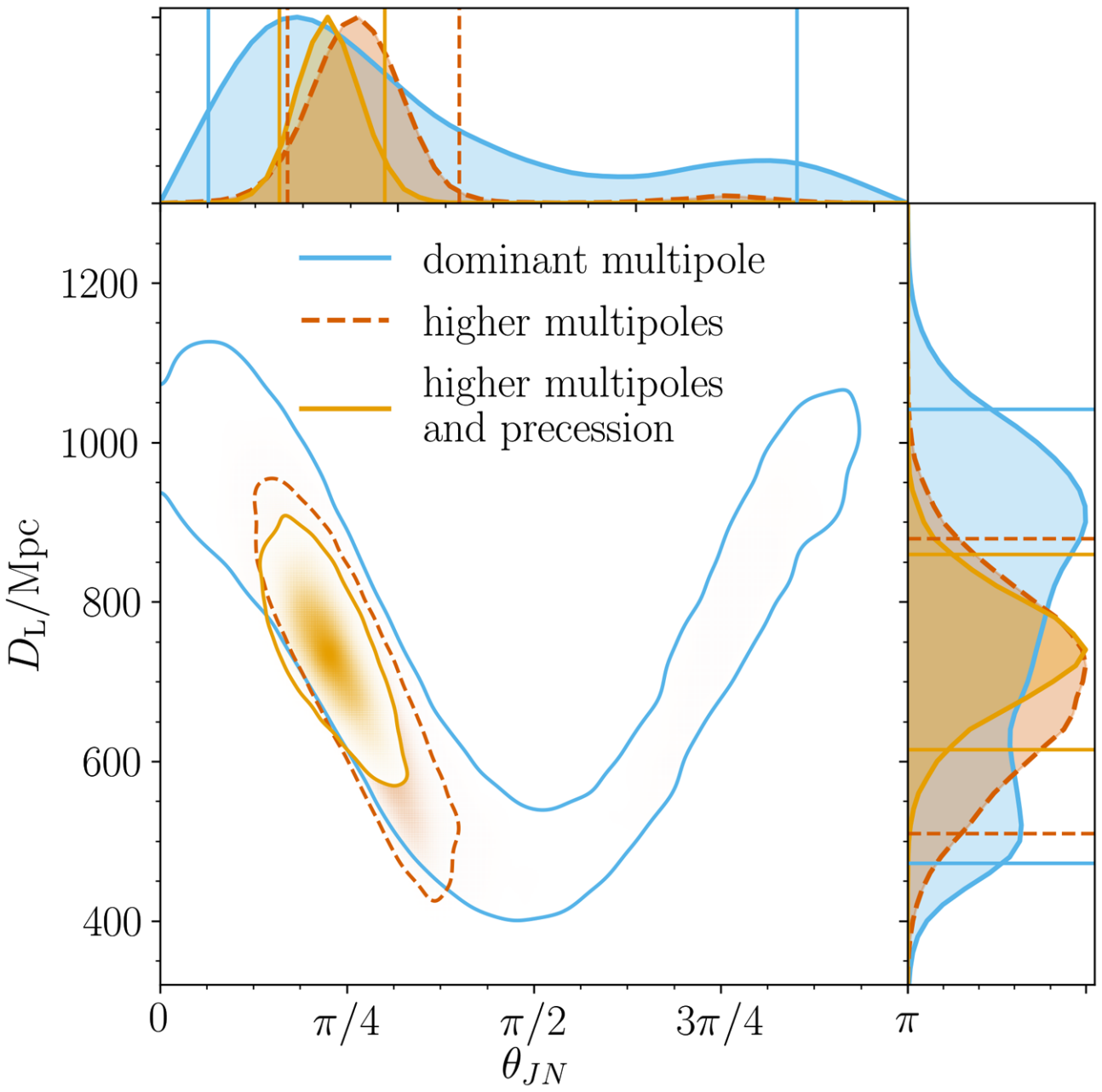

FIG. 4. The posterior distribution for the luminosity distance, $D_{\rm L}$, and inclination, $\theta_{JN}$ (angle between the line-of-sight and total angular momentum), of GW190412. We illustrate the 90% credible regions as in Fig. 2. By comparing models that include either the dominant multipole (and no precession), higher multipoles and no precession, or higher multipoles and precession, we can see the great impact higher multipoles have on constraining the inclination and distance. All models shown here are part of the EOBNR family.

We constrain the spin parameter $\chi_{\rm eff}$ of GW190412's source to be $0.25^{+0.08}_{-0.11}$. After GW151226 and GW170729 [2,7,34], this is the third BH binary we have identified whose GW signal shows imprints of at least one nonzero spin component, although recently another observation of a potentially spinning BH binary was reported [11]. However, inferred spins are more sensitive than other parameters (e.g., component masses) to the choice of the prior. A reanalysis of GW events with a population-informed spin prior recently suggested that previous binary component spin measurements may have been overestimated because of the use of an uninformative prior [137]. Collecting more observations will enable us to make more confident statements on BH spins in the future.

The parameter $\chi_{\rm eff}$ only contains information about the spin components perpendicular to the orbital plane. The in-plane spin components cause the orbital plane to precess [138], but this effect is difficult to observe, especially when the inclination angle is near 0 or $\pi$. Using models with higher multipoles, however, we constrain the inclination of GW190412 exceptionally well and put stronger constraints on the effect of precession than in previous binaries [7]. The strength of precession is parametrized by an effective precession parameter, $0 \leq \chi_{\rm p} < 1$, defined by [139]

$$\chi_{\rm p} = \max\left\{\frac{\|\vec{S}_{1\perp}\|}{m_1^2}, \kappa\frac{\|\vec{S}_{2\perp}\|}{m_2^2}\right\}, \qquad (3)$$

where $\vec{S}_{i\perp} = \vec{S}_i - (\vec{S}_i \cdot \vec{L}_N)\vec{L}_N/\|\vec{L}_N\|^2$ and $\kappa = q(4q+3)/(4+3q)$. Large values of $\chi_{\rm p}$ correspond to strong precession.

Figure 5 shows that the marginalized one-dimensional posterior of $\chi_{\rm p}$ is different from its global prior distribution. The Kullback-Leibler divergence [107], $D_{\rm KL}$, for the information gained from the global prior to the posterior is $0.95^{+0.03}_{-0.03}$ bits and $0.51^{+0.02}_{-0.02}$ bits for the EOBNR PHM and Phenom PHM model, respectively. Those values are larger than what we found for any observation during the first two observing runs (see Table V in Appendix B of [7]). Since the prior we use introduces non-negligible correlations between mass ratio, $\chi_{\rm eff}$ and $\chi_{\rm p}$, we check if the observed posterior is mainly derived from constraints on $\chi_{\rm eff}$ and $q$. We find that this is not the case, as a prior restricted to the 90% credible bounds of $q$ and $\chi_{\rm eff}$ (also included in Fig. 5) is still significantly different from the posterior, with $D_{\rm KL} = 0.98^{+0.03}_{-0.03}$ bits ($0.54^{+0.02}_{-0.02}$bits) for the EOBNR PHM (Phenom PHM) model. We constrain $\chi_p \in [0.15, 0.50]$ at 90% probability, indicating that the signal does not contain strong imprints of precession, but very small values of $\chi_p \lesssim 0.1$ are also disfavored. The results obtained with the EOBNR PHM model are more constraining than the Phenom PHM results. We return to the question if GW190412 contains significant imprints of precession below, and in the context of Bayes factors in Sec. IV A.

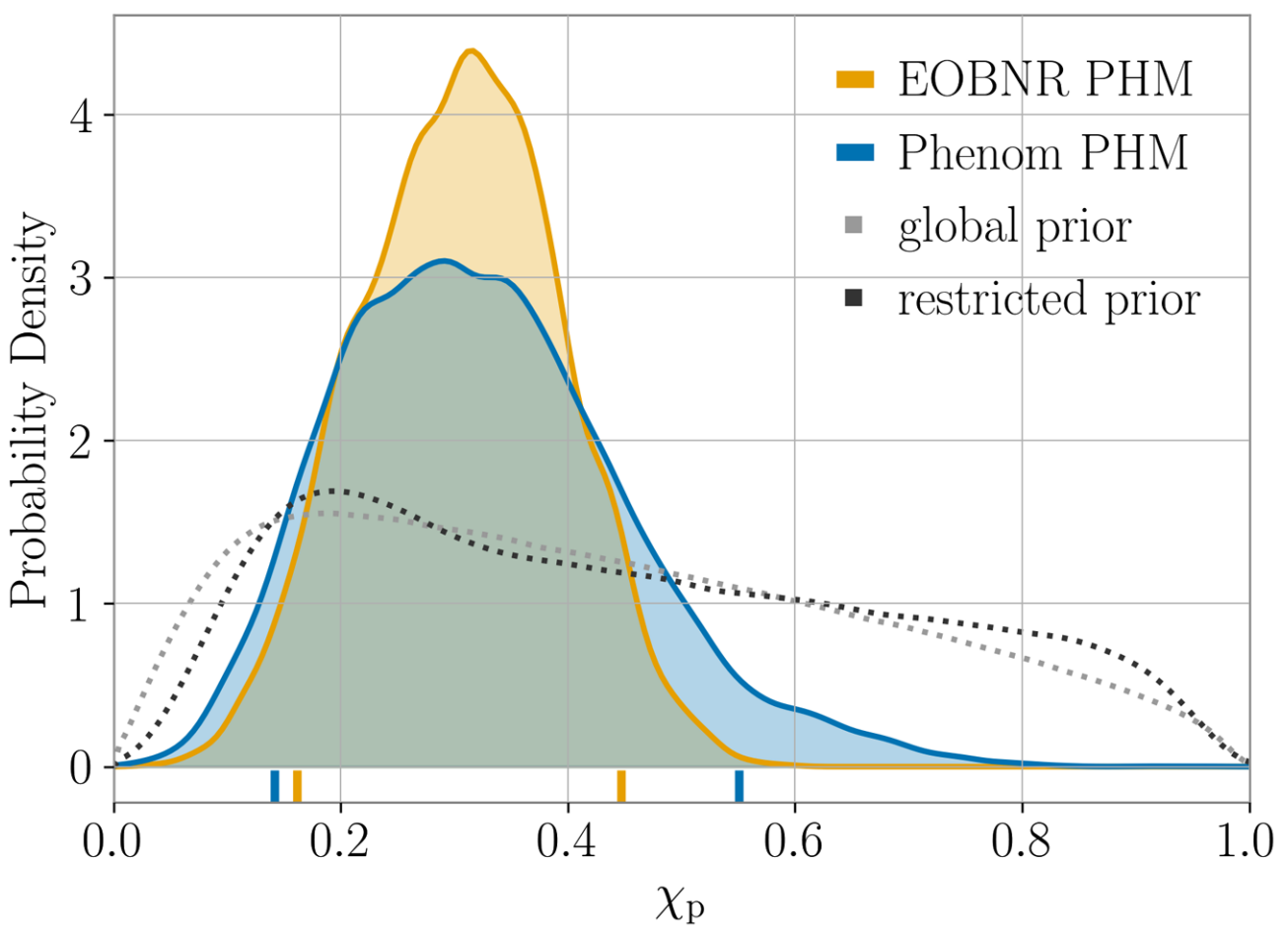

FIG. 5. The posterior density of the precessing spin parameter, $\chi_p$, obtained with the two models that include both the effects of precession and higher multipoles. In addition, we show the prior probability of $\chi_p$ for the global prior parameter space, and restricted to the 90% credible intervals of $\chi_{\rm eff}$ and $q$ as given in the "Combined" column of Table II.

Assuming a uniform prior probability between 0 and 1 for each BH's dimensionless spin magnitude, the asymmetric masses of GW190412 imply that the spin of the more massive BH dominates contributions to $\chi_{\rm eff}$ and $\chi_p$. Under these assumptions, we infer that the spin magnitude of the more massive BH is $\chi_1 = 0.44^{+0.16}_{-0.22}$, which is the strongest constraint from GWs on the individual spin magnitude of a BH in a binary so far [7]. The spin of the less massive BH remains largely unconstrained. The posterior distribution of both spin magnitudes is shown in Fig. 6. Consistent with the posterior distributions of $\chi_{\rm eff}$ and $\chi_p$, the analysis using the EOBNR PHM model constrains $\chi_1$ more than the Phenom PHM analysis.

To further explore the presence of precession in the signal, we perform the following analysis. Gravitational waveforms from precessing binaries can be decomposed into an expansion in terms of the opening angle, $\beta_{JL}$, between the total and orbital angular momenta (see Sec. III in [140], and [141]). Considering only $\ell = 2$ modes, this expansion contains five terms proportional to $\tan^k(\beta_{JL}/2)$ ($k = 0, \ldots, 4$), and each term alone does not show the characteristic phase and amplitude modulations of a precessing signal. When the spin component that lies in the binary's orbital plane is relatively small, $\beta_{JL}$ is small as well [142], and higher-order contributions in this expansion may be neglected. As a result, a precessing waveform can be modeled by the sum of the leading two contributions, where the amplitude and phase modulations of a precessing signal arise from the superposition of these terms.

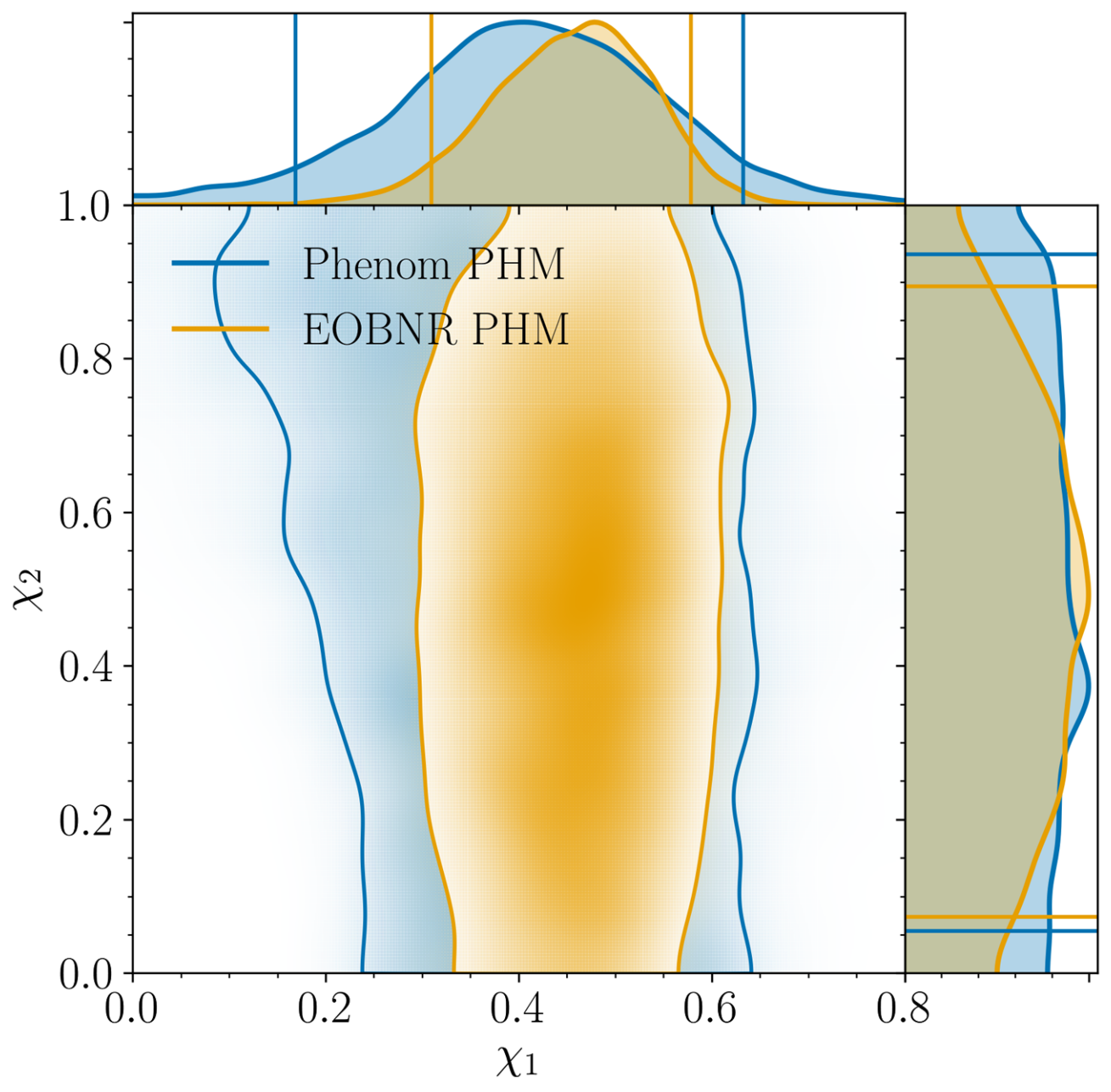

FIG. 6. The posterior distribution for the dimensionless spin magnitudes $\chi_1$ (corresponding to the more massive BH) and $\chi_2$ (less massive BH). Contours and lines indicate the 90% credible regions.

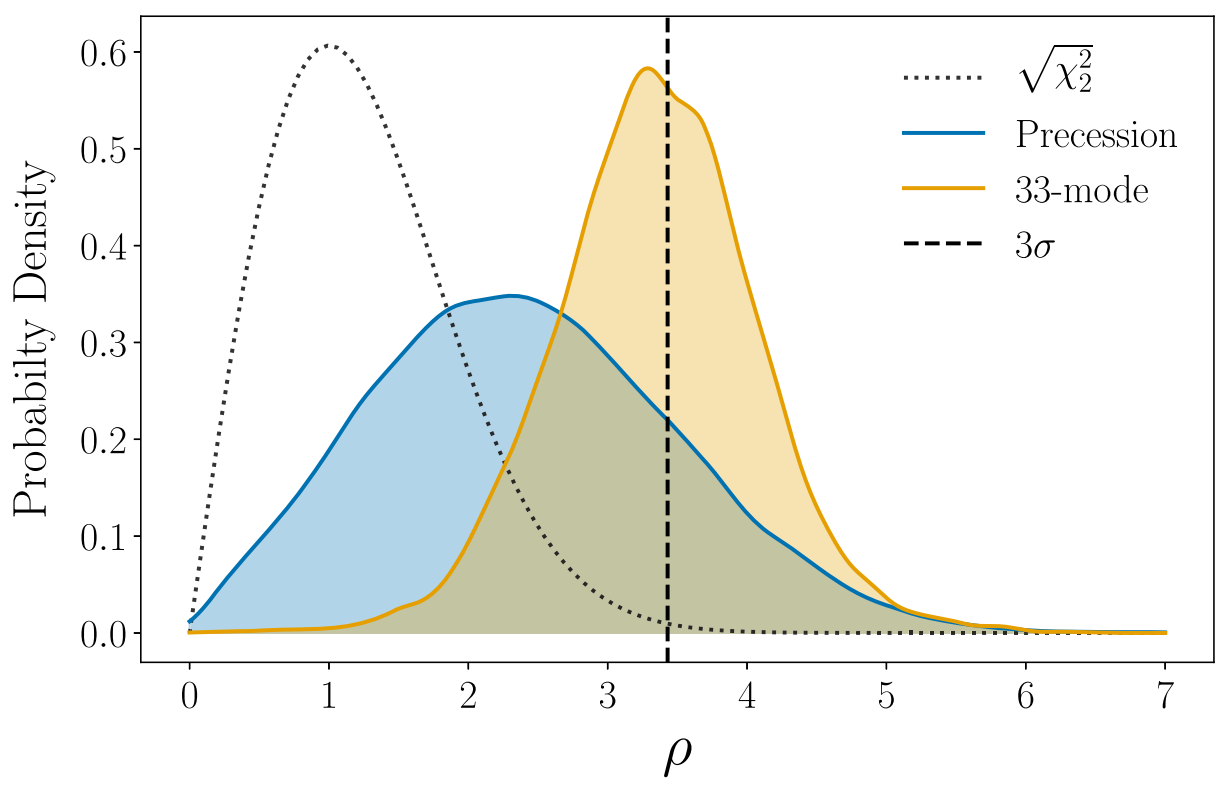

FIG. 7. The probability distribution of the precessing SNR, $\rho_p$ (blue) and the orthogonal optimal SNR, $\rho$, contained in the strongest higher multipole, $(\ell, m) = (3, 3)$ (orange). We also show the expected distribution from Gaussian noise (dotted line) and the $3 - \sigma$ level (dashed line). The results indicate that there is marginal support for precession, but the posterior supports a clearly measurable higher multipole.

In order to identify precession, we therefore require being able to measure both of these terms. We quantify the measurability of precession $\rho_{\rm p}$ by how much power there is in the subdominant contribution. The distribution of $\rho_{\rm p}$ is shown in Fig. 7. In the absence of any precession in the signal, we expect $\rho_{\rm p}^2$ to follow a $\chi^2$ distribution with 2 degrees of freedom. Using the inferred posterior distributions, our analysis shows that $\rho_{\rm p} = 2.36^{+1.96}_{-1.64}$. We may interpret this as moderate support for precession as the median exceeds the 90% confidence interval expected from noise, but a non-negligible fraction of the $\rho_{\rm p}$ posterior lies below. This calculation assumes a signal dominated by the $\ell = 2$ multipole. However, we have verified that the contribution of higher multipoles to the measurement of spin precession is subdominant by a factor of ∼5.

## IV. HIGHER MULTIPOLES

Signal models that include higher multipoles are needed to infer the strongest constraints on GW190412's source properties. This is because if the data contain significant imprints of higher multipoles, the appropriate models can fit the data better than dominant-mode models, leading to a higher statistical likelihood. Conversely, if the data would not contain imprints of higher multipoles, using more complex models allows us to disfavor configurations in which clear imprints of higher multipoles are predicted [22,135,136].

In this section, we analyze how strong the imprints of higher multipoles are in GW190412 and ask if their contributions in the data are significantly stronger than random noise fluctuations. We address this question using four different approaches, each coming with its unique set of strengths and caveats. A summary of all methods discussed below is given in Table III.

TABLE III. Summary of methods presented in Sec. IV. The statistics we report are either the Bayes factor $\mathcal{B}$ or likelihood ratio $\Lambda$ in favor of higher multipoles, or the $p$-value for the observed signal properties under a null hypothesis assuming no higher multipoles. All tests support the existence of higher multipoles in GW190412; the statistical significance varies due to the different nature of the tests.

| | |
|---|---|
| Bayes factors (Sec. IV A) | $\log_{10} \mathcal{B} > 3$ |
| Matched-filter SNR[a] (Sec. IV A) | $\log_{10} \Lambda \gtrsim 5$ |
| Overlap wavelet reconstruction[b] (Sec. IV D) | $\log_{10} \Lambda \gtrsim 1$ |
| Optimal SNR (3,3) multipole (Sec. IV B) | $p \sim 3 \times 10^{-3}$ |
| Time-frequency tracks (Sec. IV C) | $p \sim 6 \times 10^{-4}$ |

[a]Based on median difference in $\rho_{\rm HLV}$ between higher multipole and dominant multipole models (both including precession).

[b]Based on median difference in overlap using either higher multipoles or dominant multipole (nonprecessing) models; assuming $\rho_{\rm HLV} = 19$.

### A. Bayes factors and matched-filter SNR

We may first ask if higher-multipole models actually fit the data better than dominant-multipole models. This can be quantified by the matched-filter network SNR, $\rho_{\rm HLV}$, which is based on the sum of the squared inner products between the instruments' data and the signal model. Thus the SNR quantifies the extent to which a single signal model recovers coherent power between detectors. For the EOBNR family, we find that $\rho_{\rm HLV}$ increases from $18.1^{+0.2}_{-0.2}$ for the dominant-multipole model to $18.8^{+0.2}_{-0.3}$ for the higher multipole model to $19.1^{+0.2}_{-0.2}$ for the most complete EOBNR PHM model. The precessing, but dominant-multipole model yields $\rho_{\rm HLV} = 18.5^{+0.2}_{-0.3}$, which is smaller than the value for the nonprecessing, higher-multipole model. A similar trend can be observed for the Phenom family, where the dominant-multipole model, the Phenom P model, Phenom HM and Phenom PHM yield $\rho_{\rm HLV} = 18.1^{+0.2}_{-0.2}$, $18.3^{+0.2}_{-0.3}$, $18.8^{+0.2}_{-0.3}$, and $19.0^{+0.2}_{-0.3}$, respectively. The likelihood of observing the data (after maximizing over the model's overall amplitude) is proportional to $\exp(\rho_{\rm HLV}^2/2)$ [34,143]. We can therefore estimate the ratio of two likelihoods, each based on a different model, from the difference in SNR as $\Lambda = \exp(\Delta\rho_{\rm HLV}^2/2)$. Comparing the precessing, higher multipole models and their precessing, dominant multipole counterparts, we find $\Lambda$ to be between $\mathcal{O}(10^5)$ and $\mathcal{O}(10^6)$.

A more complete answer to the question of which model describes the data best can be given in the Bayesian framework. The ratio of marginalized likelihoods under two competing hypotheses is called the Bayes factor, $\mathcal{B}$ [144]. Bayes factors may be used to quantify support for one hypothesis over another. The Bayes factor does not take into account our prior belief in the hypotheses being tested. Within GR, every compact binary coalescence signal includes higher multipoles and the prior odds in favor of their presence in the signal are infinite. We therefore focus on the Bayes factors and do not discuss the odds ratio (which is the Bayes factor multiplied by the prior odds).

Table IV presents $\log_{10}\mathcal{B}$ for various combinations of two models within the same waveform family. To estimate systematic uncertainties, we test the same hypotheses using multiple model families and multiple codes to calculate $\log_{10}\mathcal{B}$. Bilby [99–101] and LALInference [102] use variants of the nested sampling algorithm [145–148]. RIFT [103,104] is based on interpolating the marginalized likelihood over a grid covering only the intrinsic source parameters. Table entries marked "…" have not been calculated because of computational constraints (LALInference analysis of precessing EOBNR models) or because the NRSurrogate model does not allow precessing spins.

We consistently find $\log_{10}\mathcal{B} \geq 3$ in favor of higher multipoles. This indicates strong evidence that the observed signal contains measurable imprints of higher multipoles, assuming either precessing or nonprecessing (aligned spin) systems. Systematic differences between codes and waveform models dominate the uncertainty in the numbers we report. We find larger differences between codes when assuming precessing spins, because this is a more complex analysis that requires exploring more degrees of freedom in the parameter space than in the nonprecessing case. However, $\log_{10}\mathcal{B}$ is large enough across all models and codes that a statement about higher multipoles can convincingly be made despite uncertainties of up to the order unity.

TABLE IV. $\log_{10}\mathcal{B}$ computed between two hypotheses that assume either a signal model including higher multipoles ($\ell \le 5$) or a dominant-multipole model ($\ell = 2$). $\log_{10}\mathcal{B} > 0$ indicates support for higher multipoles. Each entry is based on a comparison between either precessing (first row) or nonprecessing, aligned-spin (second row) models of the same model family. See Table I for full details of the models. For each family, we also indicate the code used for calculating $\log_{10}\mathcal{B}$.

| | EOBNR | | | Phenom | NRSurrogate |
|---|---|---|---|---|---|
| Hypotheses | Bilby | LALInference | RIFT | Bilby | RIFT |
| Higher vs dominant multipoles (precessing) | 4.1 | ⋯ | 3.0 | 3.6 | ⋯ |
| Higher vs dominant multipoles (aligned) | 3.5 | 3.3 | 3.6 | 3.4 | 3.4 |

It would be desirable to also compare the hypotheses that the signal contains imprints of precession with assuming no precession. However, using the same codes and models that were used in Table IV, the Bayes factors we found ranged from no decisive support for either hypothesis to indicating marginal support of precession. All values for $\log_{10}\mathcal{B}$ comparing precession vs nonprecessing models were smaller or comparable to the systematic uncertainties of order unity. More extensive studies will be needed to understand the origin of these systematics better.

### B. Optimal SNR

A complementary way to quantify the strength of higher multipoles is to use parameter-estimation results from a signal model including higher-order multipoles [149,150]. Each multipole is decomposed into parts parallel and perpendicular to the dominant multipole by calculating the noise-weighted inner product [128,151] (often referred to as overlap) between it and the dominant multipole. Among the strongest multipoles that are included in our models, the $(\ell, m) = (3, 3)$, (4,4) and (4,3) multipoles of GW190412 are close to orthogonal to the dominant (2,2) multipole within the band of the detector. In contrast, the (3,2) and (2,1) multipoles have non-negligible parallel components. To quantify the strength of the higher multipoles we remove any parallel components from the multipoles and calculate the orthogonal optimal SNR using IMRPhenomHM [22]. We find $(\ell, m) = (3, 3)$ to be the strongest subdominant multipole.

The templates that include higher multipoles do not allow the amplitude and phase of the (3,3) multipole to be free parameters; they are determined by the properties of the system. An analysis of this event using only the dominant (2,2) multipole recovers posteriors that are consistent with a broad range of inclinations, coalescence phases, and mass ratios, while the same analysis using higher multipoles results in significantly more restricted posteriors (see Fig. 4). This suggests that by changing those parameters, our models can effectively treat the amplitude and phase of the higher multipoles as tunable parameters that make their contributions more or less pronounced. If the data only contained the dominant quadrupole mode and Gaussian noise, the squared orthogonal SNR in the subdominant multipole will be $\chi^2$ distributed with 2 degrees of freedom [140,141,150]. This was verified by analyzing an injection with parameters close to GW190412.

This noise-only distribution is shown in Fig. 7, along with the orthogonal optimal SNR in the $(\ell, m) = (3, 3)$ mode. The peak of the SNR distribution is at the Gaussian equivalent $3-\sigma$ level for the noise-only distribution (i.e., with cumulative tail probability of $p = 3 \times 10^{-3}$), making this the most significant evidence for something other than the dominant multipole to date [152].

### C. Time-frequency tracks

An independent analysis was performed to detect the presence of higher-order multipoles in the inspiral part of the signal, using the time-frequency spectrum of the data. Full details of the approach are described in [153], but we summarize the main idea and results for GW190412 below.

The instantaneous frequency $f_{\ell m}(t)$ of the GW signal from an inspiraling compact binary is related to that of the dominant (2,2) mode by $f_{\ell m}(t) \simeq (m/2) f_{22}(t)$. We compute $f_{22}(t)$ from the dominant multipoles of the EOBNR HM and Phenom HM models, using the maximum likelihood source parameters from the standard analysis presented in Sec. III. Inspired by the above scaling relation, we then look along the generalized tracks, $f_\alpha(t) \simeq \alpha f_{22}(t)$, in a time-frequency representation that is the absolute square of the continuous wavelet transformation (CWT) of the whitened on-source data, $\tilde{X}(t, f)$. We have used Morlet wavelets to perform the CWT, where the central frequency of the wavelet was chosen so as to maximize the sum of pixel values along the $f_{22}(t)$ curve. This wavelet transformation is shown in the top panel of Fig. 8.

In order to quantify the energy along each track, we define $Y(\alpha)$ to be the energy $|\tilde{X}(t, f)|^2$ in all pixels containing the track $f_\alpha(t)$, where we discretize the data with a pixel size of $\Delta t = 1/4096$ s along the time axis and

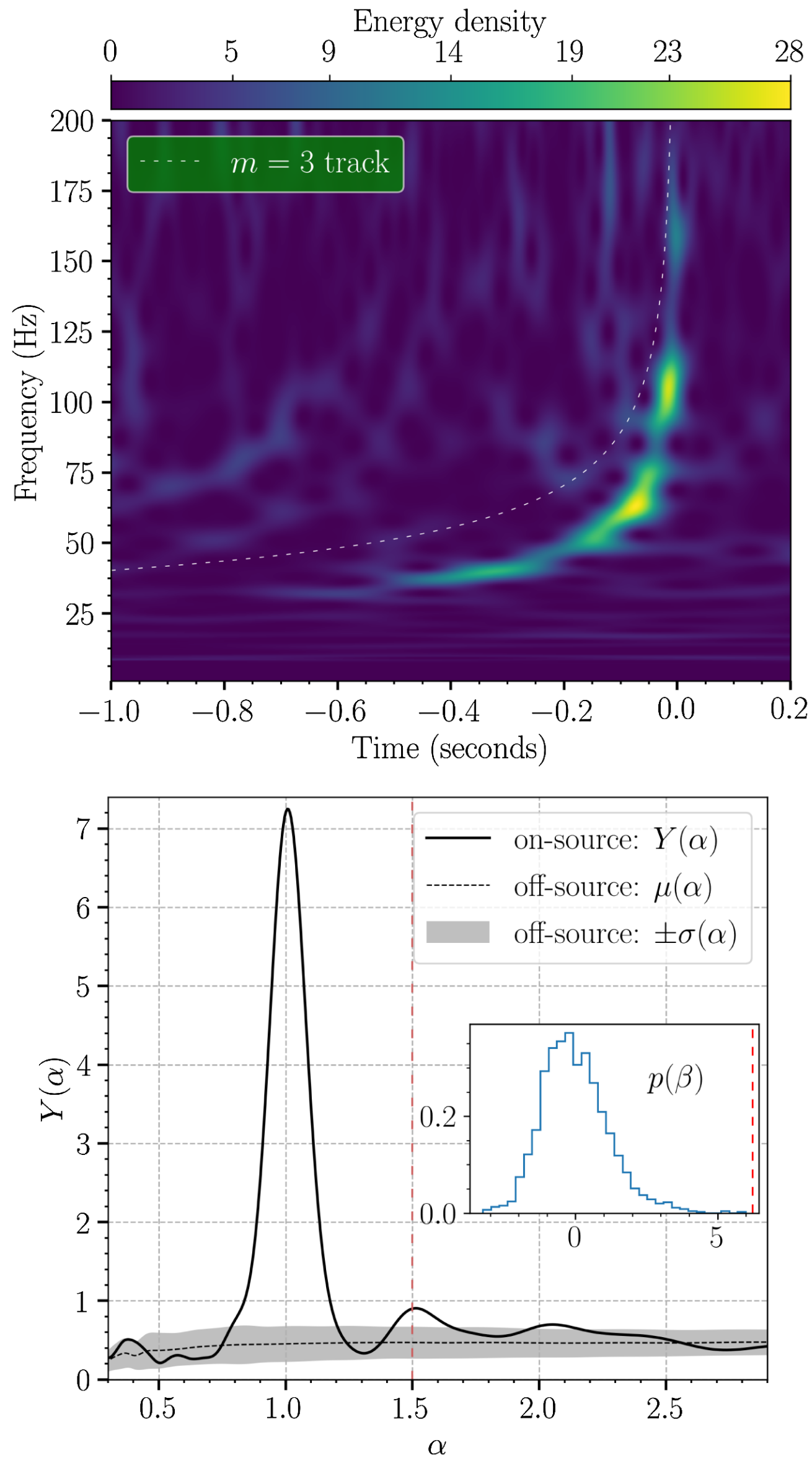

FIG. 8. Top panel: Time-frequency spectrogram of data containing GW190412, observed in the LIGO Livingston detector. The horizontal axis is time (in seconds) relative to the trigger time (1239082262.17). The amplitude scale of the detector output is normalized by the PSD of the noise. To illustrate the method, the predicted track for the $m = 3$ multipoles is highlighted as a dashed line, above the track from the $m = 2$ multipoles that are visible in the spectrogram. Bottom panel: The variation of $Y(\alpha)$, i.e., the energy in the pixels of the top panel, along the track defined by $f_\alpha(t) = \alpha f_{22}(t)$, where $f_{22}(t)$ is computed from the Phenom HM analysis. Two consecutive peaks at $\alpha = 1.0$ and $\alpha = 1.5$ (thin dashed line) indicate the energy of the $m = 2$ and $m = 3$ multipoles, respectively. Inset: The distribution of the detection statistic $\beta$ in noise, used to quantify $p$-values for the hypothesis that the data contains $m = 2$ and $m = 3$ multipoles (red dashed line).

$\Delta f = 1/5$ Hz along the frequency axis. By doing so, we can decouple the energy in individual multipoles of the signal. Once $f_{22}(t)$ is defined, this is a computationally efficient way to analyze which multipoles have sufficient energy to be detectable in the data; no further modeling input is needed, although we do not require phase coherence along each track.

The resulting $Y(\alpha)$ for GW190412 is shown in the bottom panel of Fig. 8. It has a global peak at $\alpha = 1$, corresponding to the dominant (2,2) multipole, and a prominent local peak at $\alpha = 1.5$, corresponding to the $m = 3$ multipoles. We also calculate $Y(\alpha)$ from different segments surrounding, but not including GW190412, to capture the detector noise characteristics; in this case we call the quantity $N(\alpha)$. The ensemble average $\mu(\alpha) = \langle N(\alpha) \rangle$ and standard deviation $\sigma(\alpha)$ of $N(\alpha)$ are also plotted for reference and to highlight the relative strength of the GW signal present in the on-source segment.

Instead of estimating the significance of the $m = 3$ multipoles from comparing $Y(\alpha)$ to its background at $\alpha = 1.5$, we perform a more powerful statistical analysis in which we test the hypotheses that the data contain either only noise ($\mathcal{H}_0$), or noise and a dominant-multipole maximum likelihood signal ($\mathcal{H}_1$), or noise and a maximum likelihood signal that includes $m = 2$ and $m = 3$ multipoles ($\mathcal{H}_2$). By maximizing the likelihood of observing $Y(\alpha)$ given each hypothesis over a free amplitude parameter for each multipole, we obtain likelihood ratios for $\mathcal{H}_1$ and $\mathcal{H}_2$, and their difference is in turn incorporated into a detection statistic $\beta$ [see Eq. (7) in [153] ].

From the on-source data segment taken from the LIGO Livingston detector, we found the detection statistic $\beta = 6.1$ with a $p$-value of $6.4 \times 10^{-4}$ for EOBNR HM model; and $\beta = 6.2$ with a $p$-value $< 6.4 \times 10^{-4}$ for Phenom HM model, which strongly supports the presence of $m = 3$ modes in the signal. The full distribution of $\beta$ from off-source data segments from the LIGO Livingston detector surrounding the trigger time of GW190412 is shown in the inset of Fig. 8.

### D. Signal reconstructions

As an additional test of consistency, and an instructive visual representation of the observed GW signal, we compare the results of two signal reconstruction methods. One is derived from the parameter-estimation analysis presented in Sec. III, the second uses the model-agnostic wavelet-based burst analysis BayesWave [154] which was also used to generate PSDs. A detailed discussion of such signal comparisons for previous BBH observations can be found in [155].

For GW190412, both signal reconstruction methods agree reasonably well as illustrated in Fig. 9. To quantify the agreement for each signal model from the Phenom and EOB families, we compute the noise-weighted inner product [128,151] between 200 parameter-estimation samples and the BayesWave median waveform. The BayesWave waveform is constructed by computing the median values at every time step across samples. Similar comparison strategies have been used in [3,5,34,156].

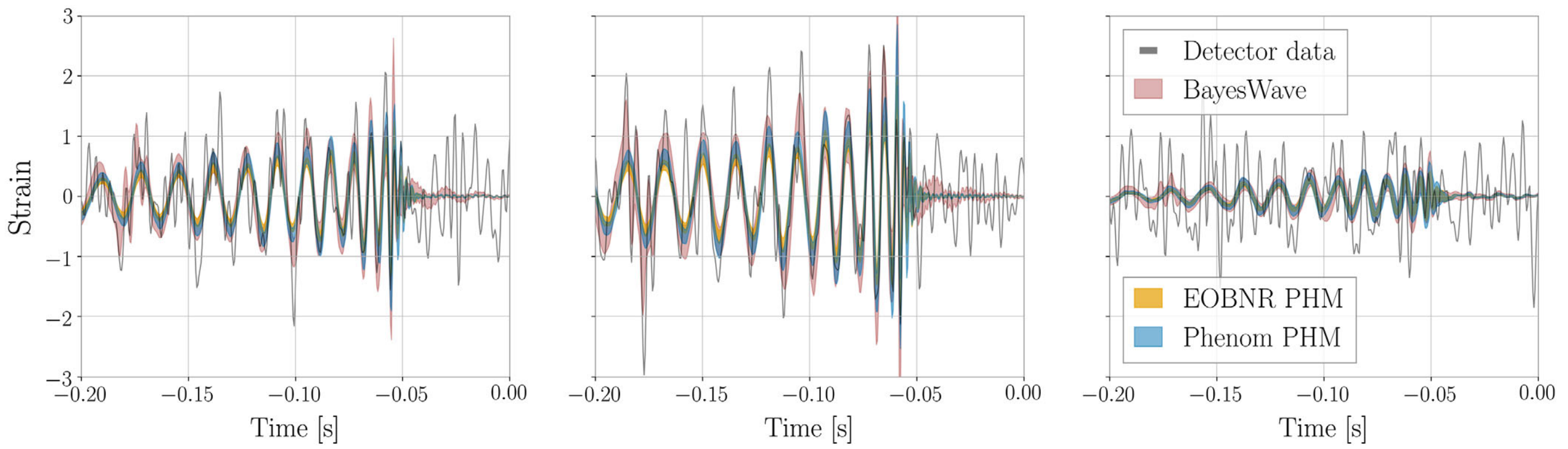

FIG. 9. Reconstructions of the gravitational waveform of GW190412 in the LIGO Hanford, LIGO Livingston and Virgo detectors (from left to right). We show detector data, whitened by an inverse amplitude-spectral-density filter computed using BayesLine [105], together with the unmodeled BayesWave reconstruction that uses a wavelet bases, and the reconstruction based on the precessing, higher multipole models from the EOBNR and Phenom families. The bands indicate the 90% credible intervals at each time. We caution that some apparent amplitude fluctuations in this figure are an artifact of the whitening procedure.

For the models of the EOBNR family, we find that the agreement with the unmodeled BayesWave reconstruction increases slightly from overlaps of $0.84^{+0.01}_{-0.02}$ for the dominant-multipole, nonprecessing model to $0.86^{+0.01}_{-0.02}$ when higher multipoles are included, to $0.88^{+0.01}_{-0.02}$ for the most complete EOBNR PHM model (median overlap and 90% errors). Increasing overlaps are consistent with the findings of the other methods presented in this section that indicate that the extra physical effects included in the higher-multipole precessing model match the data better. The overlaps we find are also consistent with expectations from [155]. The Phenom family may suggest a similar trend. The overlap for nonprecessing dominant multipole model is $0.84^{+0.02}_{-0.02}$, and it increases to $0.85^{+0.01}_{-0.03}$ for the nonprecessing higher multipoles model, to $0.86^{+0.02}_{-0.02}$ for the precessing higher multipoles model Phenom PHM. To compare those values to other methods, we relate the difference in the square overlap $\Delta O^2$ to the likelihood ratio through $\Lambda = \exp(\rho^2_{\rm HLV}\Delta O^2/2)$ [143] and find $\Lambda$ between $\mathcal{O}(10)$ and $\mathcal{O}(100)$ in favor of the higher multipole model.

## V. TESTS OF GENERAL RELATIVITY

As the first detected BBH signal with a mass ratio significantly different from unity, GW190412 provides the opportunity to test GR in a previously unexplored regime. Due to the mass asymmetry, this signal contains information about the odd parity multipole moments. Hence the tests of GR reported here are sensitive to potential deviations of the multipolar structure away from GR [157]. A violation from GR may arise from how the signal is generated by the source; additionally, the form of the signal described by GR may be tested by checking the consistency of independently obtained estimates of parameters between the inspiral and merger-ringdown parts of the full BBH waveform. The following analyses are done by using the LALInference library [102] to generate posterior probability distributions on these parameters by using the nested sampling algorithm.

### A. Constraints on gravitational wave generation

We check the consistency of this source with general relativistic source dynamics by allowing for parametrized deformations in each phasing coefficient in the binary's waveform. They were first performed on inspiral-only waveforms in [158,159] and an extension to higher-modes was studied in [160]. The current version of the test using the phenomenological waveform models rely on extensions of the methods laid out in [161,162]. Such tests have been performed on all GW detections made in O1 and O2 [163,164] and have been updated recently with the best constraints by combining all significant BBH detections made during O1 and O2 in [165]. We perform this analysis with the precessing, dominant-multipole phenomenological model Phenom P, and, for consistency with previous tests, with the aligned-spin dominant-mode EOBNR model (see Table I).

The inspiral regime of both waveforms is modeled using the PN approximation. The fractional deviation parameters $\delta\hat{\varphi}_n$ are added to their respective PN coefficients $\varphi_n$ at $n/2$-PN order. While the deviation coefficients are added differently in the two models, the differences are taken care of by effectively reparametrizing the coefficients added to the EOB-based model for consistency in comparing bounds from the Phenom-based model. The full set of parameters being tested is $\{\delta\hat{\varphi}_0, \delta\hat{\varphi}_1, \delta\hat{\varphi}_2, \delta\hat{\varphi}_3, \delta\hat{\varphi}_4, \delta\hat{\varphi}_{5l}, \delta\hat{\varphi}_6, \delta\hat{\varphi}_{6l}, \delta\hat{\varphi}_7\}$. Here, $\delta\hat{\varphi}_{5l}$ and $\delta\hat{\varphi}_{6l}$ refer to the fractional deviations added to the log-dependent terms at 2.5PN and 3PN respectively. Moreover, $\delta\hat{\varphi}_1$ is an absolute deviation as there is no 0.5PN term within GR. These parameters are tested by varying only one $\delta\hat{\varphi}_n$ at a time, and introducing these to the parameter set of the full signal model. This increased parameter space dimensionality makes it especially challenging to use the already computationally expensive Phenom PHM waveform, and we restrict our analysis to using the precessing Phenom P approximant. To check for possible systematics introduced from analyzing this signal with the dominant-mode Phenom P model, a signal similar to GW190412 was created with the Phenom PHM waveform model and injected into data generated using the BayesWave PSDs for the event. The recovery of this signal, using Phenom P shows that the posteriors are completely consistent with the injected values, suggesting that for a GW190412-like signal with the same SNR, the inclusion of the higher multipoles does not significantly bias results when those multipoles are not included.

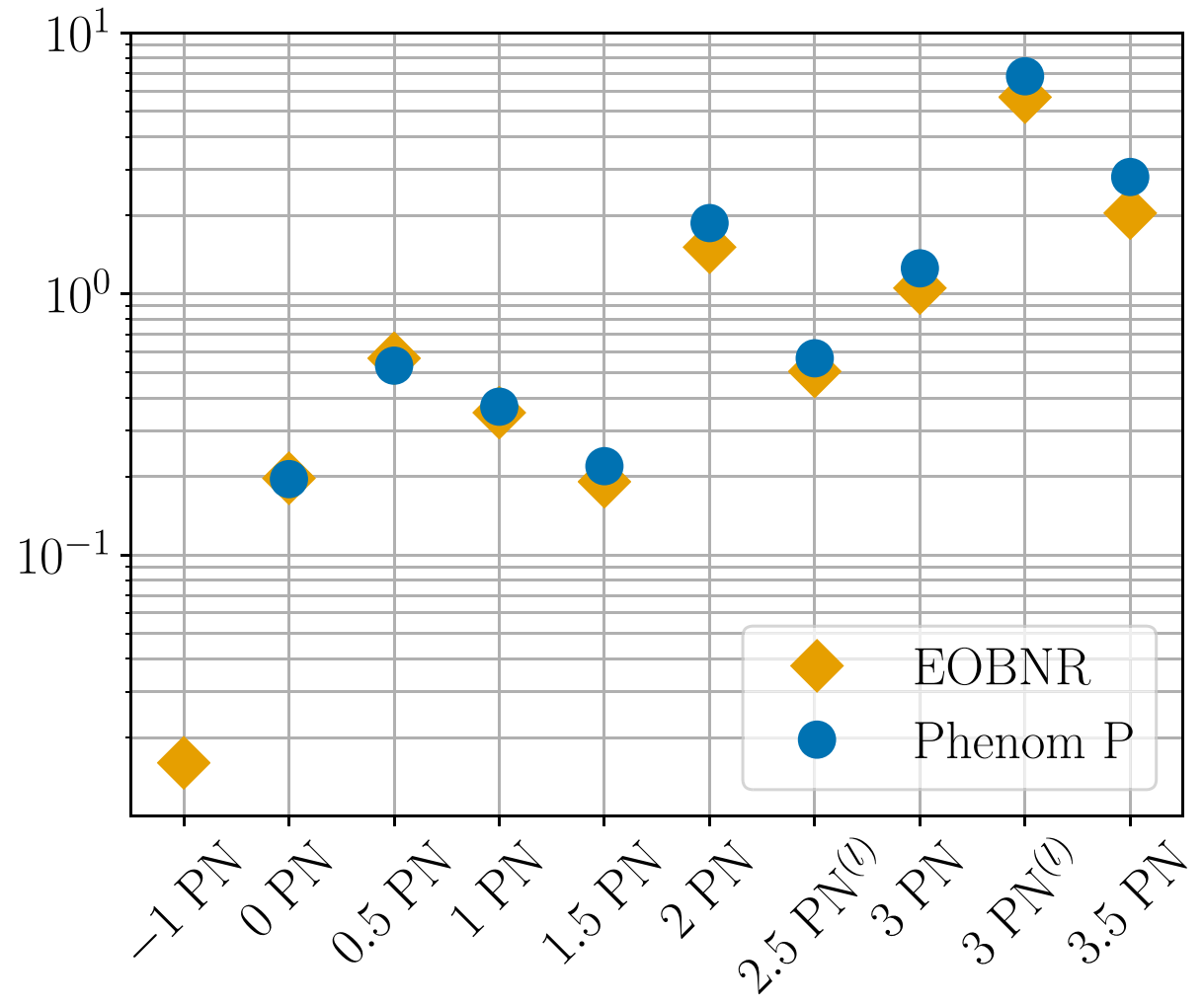

FIG. 10. 90% upper bounds obtained from parametrized deviations in PN coefficients.

The posterior distributions on the fractional deviation parameters are always found to be consistent with the GR prediction of $\delta\hat{\varphi}_n = 0$. Additionally, the EOB model can test deviations in the $-1$ PN dipole term, while the phenomenological signal model can be used to test the intermediate and merger-ringdown parameters of the signal, which are consistent with their GR value. However, owing to the longer inspiral of this signal, the bounds obtained from this event in the inspiral regime are among the most constraining bounds obtained from the analyses on individual BBH detections as reported in [165]. We show the 90% upper bounds computed in Fig. 10. The only BBH signals that give more robust or comparable bounds above 0 PN order are those from GW170608 (the lowest mass BBH to have been published) and GW151226.

### B. Inspiral-merger-ringdown consistency

We check the consistency of signal parameters between the low- and high-frequency parts of the signal [166,167]. Estimates of the final mass $M_\mathrm{f}$ and final spin $\chi_\mathrm{f}$ of the remnant BH are found from the two parts of the frequency domain signal and their fractional differences are checked for consistency. For this source, the transition from the lower-frequency to higher-frequency part (the Kerr innermost stable circular orbit) of the signal is estimated from the median intrinsic source parameters and the resulting prediction for $M_\mathrm{f}$ and $\chi_\mathrm{f}$ to be at $f = 211$ Hz [168]. We used the signal model Phenom PHM, sampling on the BBH parameter set to obtain posterior probability distributions on all parameters. The component masses and spins are estimated directly from the lower-frequency part of the signal, and, using numerical relativity fit formulas [124–127], the posteriors on $M_\mathrm{f}$ and $\chi_\mathrm{f}$ are inferred. Despite the weak constraints on the spin magnitude of the less massive BH, the final spin is well constrained as it is dominated by the binary's orbital angular momentum and the total spin angular momentum, which in turn is the mass-squared weighted sum of $\chi_1$ and $\chi_2$.

From the higher-frequency part of the signal, estimates on component masses and spins are obtained again using the same waveform model, and the posteriors on $M_\mathrm{f}$ and $\chi_\mathrm{f}$ are inferred using the same fit formulas as above. From those two distributions, a posterior distribution of the fractional differences, denoted by $\Delta M_\mathrm{f}/\bar{M}_\mathrm{f}$ and $\Delta\chi_\mathrm{f}/\bar{\chi}_\mathrm{f}$ respectively, is then computed. Here, $\bar{M}_\mathrm{f}$ and $\bar{\chi}_\mathrm{f}$ denote the mean values of the distributions of $M_\mathrm{f}$ and $\chi_\mathrm{f}$ respectively. While we expect mass and spin differences of exactly (0,0) given a pure signal obeying GR, the presence of detector

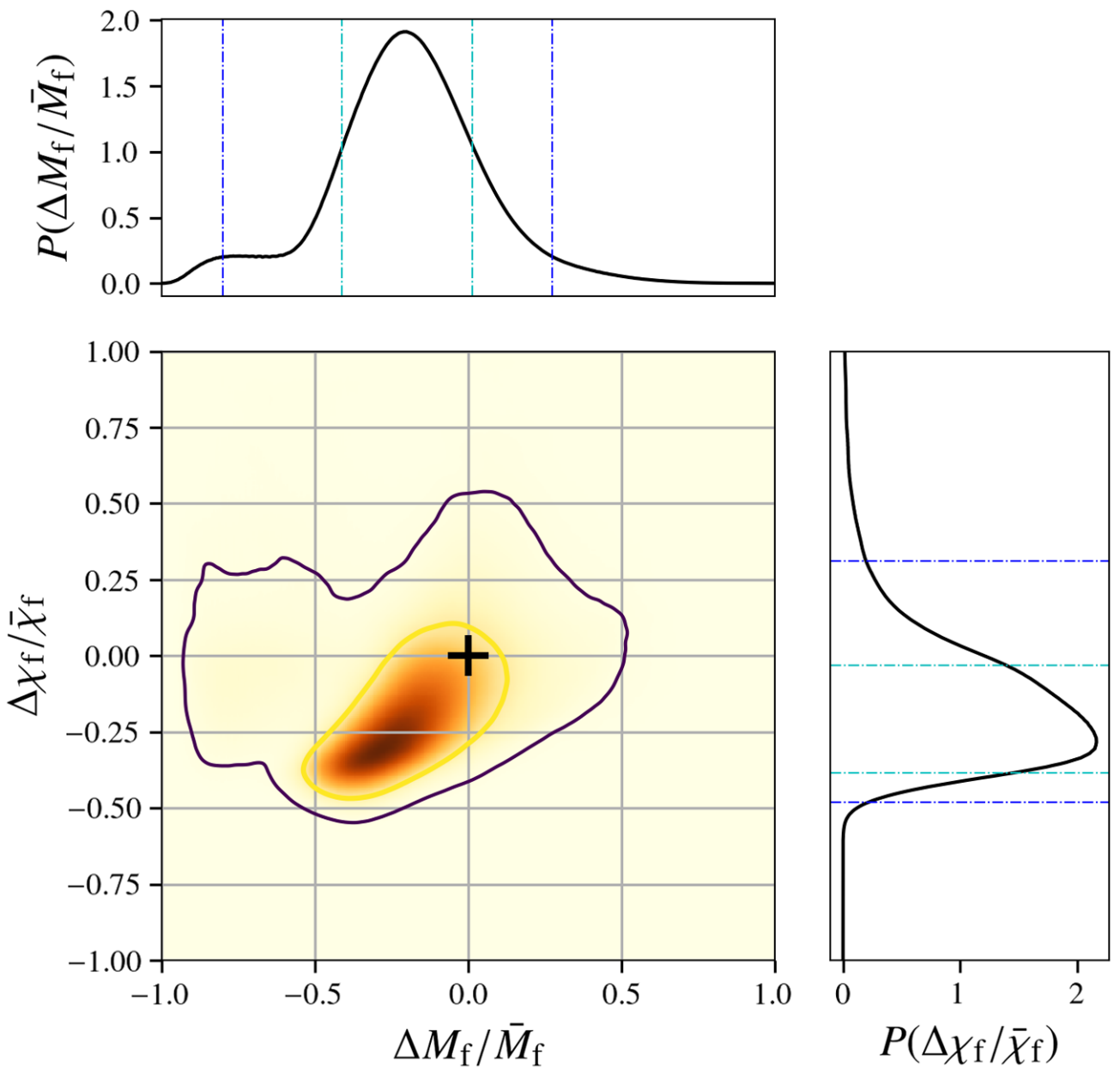

FIG. 11. Posteriors on fractional differences of GW190412's final mass and final spin inferred from either the low-frequency or high-frequency part of the signal. The GR value of 0 for both the parameters is marked by "+." The brown contour encloses 90% probability and the yellow contour encloses 68% probability.

noise will generally yield a posterior with some nonzero spread around (0,0).

Figure 11 shows the results of the posterior distributions on these fractional quantities. The 68% credible regions of the quantities $\Delta M_\mathrm{f}/\bar{M}_\mathrm{f}$ and $\Delta\chi_\mathrm{f}/\bar{\chi}_\mathrm{f}$ enclose (0,0) as can be seen from both the one-dimensional posteriors and the two-dimensional contours. GW190412 results are consistent with past observations of BBHs [165].

## VI. IMPLICATIONS FOR BBH POPULATION PROPERTIES

BBHs detected by the LIGO-Virgo network can be used to constrain the uncertain physical processes inherent to compact binary formation channels. As the first observed BBH with definitively asymmetric masses, the inclusion of GW190412 in the current catalog of BBHs has a significant impact on inferred population properties. Here, we examine (i) how the addition of GW190412 to the catalog of BBHs from the first and second observing runs affects population statements; (ii) the robustness of the component mass measurements of GW190412 when evaluated as part of the previously observed population; and (iii) whether this system's mass ratio is a significant outlier with respect to that population.

Using the ten significant BBH events in the catalog of GWs from the first and second observing runs (GWTC-1, [7]), we constrained the parameters of phenomenological models that represent the underlying BBH population [15]. In certain models, the mass-ratio distribution is parametrized with a power law, $p(m_2|m_1) \propto q^{\beta_q}$ [169–171], where $\beta_q$ is the spectral index of the mass ratio distribution. Since all ten events from GWTC-1 are consistent with symmetric masses, the posteriors for $\beta_q$ showed a preference for positive values [15], providing initial evidence supporting equal-mass pairings over randomly drawn mass pairings [172]. However, the steepness of $\beta_q$ was unconstrained, which limited our ability to determine how prevalent equal-mass pairings are relative to their asymmetric counterparts. The inclusion of GW190412 in the population provides a much stronger constraint on the mass ratio spectral index, as shown in Fig 12. Applying the population of significant events from the first and second observing runs as well as GW190412 to the simplest model that invokes a power-law spectrum to the mass ratio distribution (model B from [15]), we find $\beta_q < 2.7(5.8)$ at the 90% (99%) credible level. This indicates that though equal-mass pairings may still be preferred, there is significant support for asymmetric mass pairings; the posterior population distribution [15,173] for this model indicates that $\gtrsim 10\%$ of merging BBHs should have a mass ratio of $q \lesssim 0.40$. In fact, the support for $\beta_\mathrm{q} \leq 0$ in the recovered distribution indicates that the true mass ratio distribution may be flat or even favor unequal-mass pairings. This is not in tension with the mass ratios of the already-observed population; though all mass ratio posteriors from GWTC-1 are consistent with $q = 1$, they also have significant support for lower values. These constraints on $\beta_q$ are preliminary and final results from O3 will only be obtained after analyzing the population that includes all BBH events from this observing run.

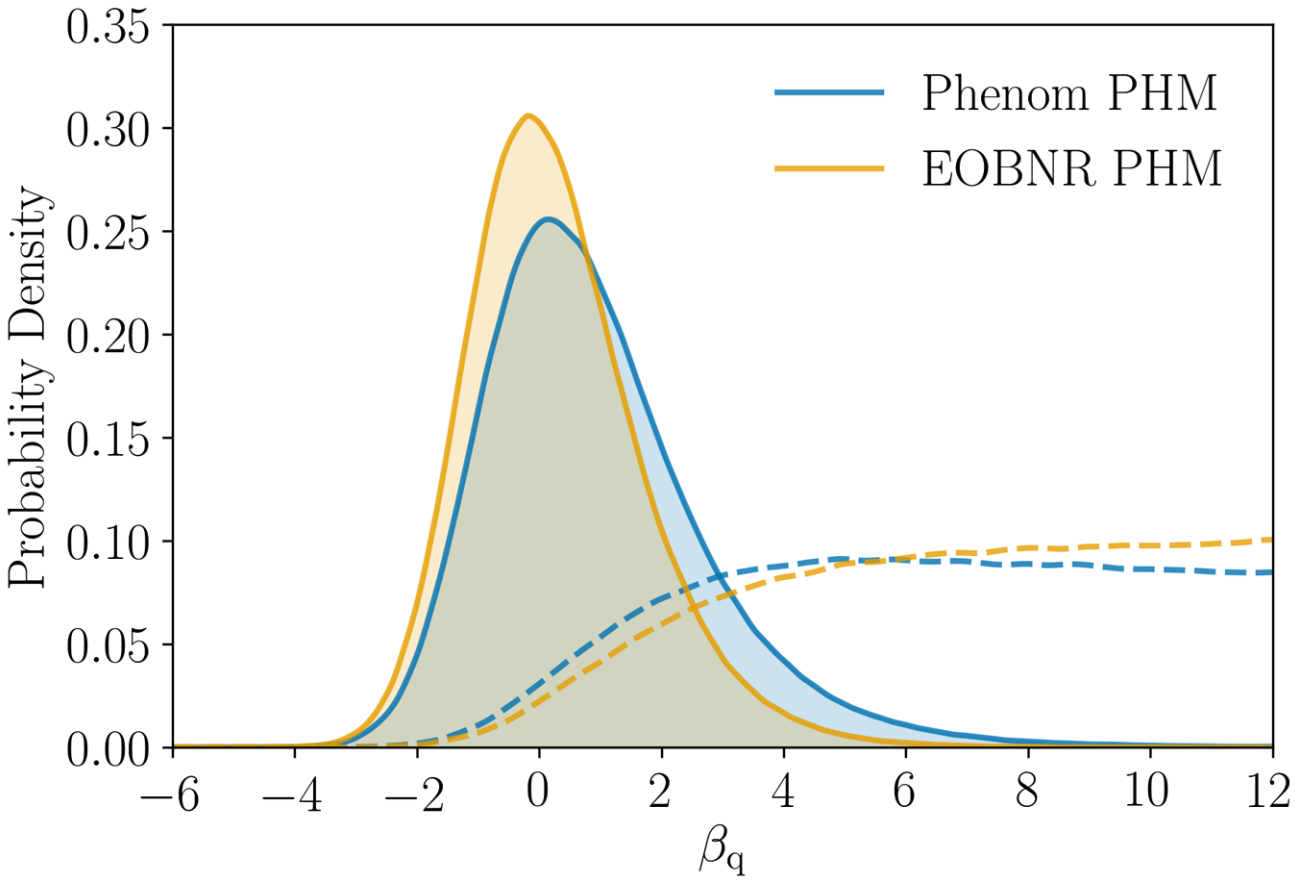

FIG. 12. Posterior on mass ratio spectral index $\beta_\mathrm{q}$ with (solid lines) and without (dashed lines) the inclusion of GW190412. We show inference using both the EOBNR and Phenom families; for the ten events from GWTC-1 we use the publicly available samples for both of these waveform families presented in [7], and for GW190412 we use the EOBNR PHM and Phenom PHM posterior samples presented in this paper.

We also check whether the asymmetric mass ratio for GW190412 is robust when the component mass posterior distributions are reweighted using a population-informed prior based on model B from [15]. Since the majority of observed systems are consistent with equal mass, the mass ratio posterior for GW190412 pushes closer towards equal mass when using a population-informed prior rather than the standard uninformative priors used to generate posterior samples. However, the mass ratio of GW190412 is still constrained to be $q < 0.43(0.59)$ at the 90% (99%) credible level, compared to $q < 0.37(0.48)$ using the standard priors from parameter estimation.

Using methods from [174], we test the consistency of GW190412 with the population of BBHs inferred from the first and second observing runs. We first construct a population model (model B) derived only from the events in GWTC-1, following the prescription in [15], and draw 11 observations from this model (representing the ten significant BBHs from the first two observing runs as well as GW190412). Examining the lowest mass ratio drawn from each set of $10^7$ such realizations, we find the population-weighted mass ratio posterior samples of GW190412 lie at the $1.7^{+10.3}_{-1.3}$ percentile of the cumulative distribution of lowest mass ratios. This indicates that given the BBH population properties inferred from the first two observing runs, drawing a system with a mass ratio analogous to GW190412 would be relatively rare. The apparent extremity of GW190412 is likely driven by the limited observational sample of BBHs and the lack of constraining power on the mass ratio spectrum prior to the observation of GW190412. Constructing a population model that includes the observation of GW190412 in the fit, we find GW190412 to lie at the $25^{+47}_{-17}$ percentile of the cumulative distribution of lowest mass ratios, indicating that GW190412 is a reasonable draw from the updated population.

## VII. ASTROPHYSICAL FORMATION CHANNELS FOR GW190412

Multiple astrophysical channels are predicted to produce the merging BBHs identified by the LIGO-Virgo network. The majority of these channels have mass ratio distributions that peak near unity, but also often predict a broad tail in the distribution that extends to more asymmetric masses. Though a wide array of formation channels exist, each with distinct predictions for merger rates and distributions of intrinsic BBH parameters, most channels can be broadly categorized as the outcome of either isolated binary stellar evolution or dynamical assembly (for reviews, see [175,176], respectively).

In the canonical isolated binary evolution scenario, by which a compact binary progenitor achieves a tight orbital configuration via a common envelope phase [175,177–182], BBH mergers with mass ratios of $q \lesssim 0.5$ are typically found to be less common than their near-equal-mass counterparts by an order of magnitude or more [183–188], though certain population synthesis modeling finds BBH mergers with asymmetric masses to be more prevalent [189,190]. However, even if the formation probability of asymmetric mass ratio systems is an order of magnitude lower than the formation probability of equal mass systems, the observation of one clearly asymmetric mass system given the current observational catalog is unsurprising. In contrast, the progenitor of GW190412 is unlikely to have formed through chemically homogeneous evolution, as this scenario typically cannot form binaries with mass ratios below $q < 0.5$ [191–193]. The asymmetric component masses of GW190412 may also suggest formation in an environment with lower metallicity, as lower metallicities are predicted to produce a higher rate of merging BBHs having mass ratios consistent with GW190412 [182,183,187,188], though this prediction is not ubiquitous across population synthesis models [194].

Dense stellar environments, such as globular clusters [195–198], nuclear clusters [199], and young open star clusters [200–206], are also predicted to facilitate stellar-mass BBH mergers. Numerical modeling of massive and dense clusters suggests that significantly asymmetric component masses are strongly disfavored for mergers involving two first-generation BHs that have not undergone prior BBH mergers (e.g., [207]). However, asymmetric component masses in dense clusters may be attained by a first-generation BH merging with a higher-generation BH that has already undergone a prior merger or mergers [170,208,209]. For formation environments such as globular clusters, this would require low natal spins for the initial population of BHs so that an appreciable number of merger products can be retained in the shallow gravitational potential of the cluster [209,210]. On the other hand, $N$-body simulations of open star clusters recover a flatter distribution in BBH mass ratio for their (first-generation) merger population [204], and can even find highly asymmetric mergers to be more prevalent than equal-mass mergers for steeper slopes of the initial BH mass spectrum [203]. BBH mergers with asymmetric masses may also be the result of massive-star collisions in young star clusters, as this physical process has been shown to amplify unequal-mass BBH mergers relative to their isolated counterparts [205].

Other formation scenarios may also be efficient at generating BBH mergers with significantly asymmetric component masses. BBHs in triple or quadruple systems can undergo Lidov-Kozai oscillations [211,212] that may expedite the GW inspiral of the binary. Such systems can either exist in the galactic field with a stellar-mass outer perturber [213–219] or in galactic nuclei with a supermassive BH as the tertiary component [220–223]. Though most modeling of hierarchical stellar systems does not include robust predictions for mass ratio distributions of merging BBHs, certain models find galactic field triples with asymmetric masses for the inner BBH to have a merger fraction that is about twice as large as their equal-mass counterparts [213]. In the context of hierarchical triples in galactic nuclei, recent modeling predicts a significant tail in the mass ratio distribution of merging BBHs that extends out to mass ratios of $\sim 8:1$ [223]. BBH mergers with significantly asymmetric component masses are also predicted for systems formed in the disks of active galactic nuclei [224–230]. The deep gravitational potential near the vicinity of the supermassive BH may allow for stellar-mass BHs to go through many successive mergers without being ejected by the relativistic recoil kick, leading to BBH mergers with highly asymmetric masses [229,230]. Though these channels may not be dominant, they could explain a fraction of the sources observed by the LIGO-Virgo network.

Spins also encode information about the formation mechanisms of BBHs. The effective spin of GW190412 is consistent with being nonzero at the 90% credible level (see Fig. 2), indicating that $\vec{S} \cdot \vec{L}_N > 0$ for at least one of the BH components. Predictions for the spins of BHs at birth are highly uncertain, and are dependent on the efficiency of angular momentum transport in their massive-star progenitors as well as prior binary interactions, e.g., [231]. Recent work modeling the core-envelope interaction in massive stars finds angular momentum transport to be highly efficient, leading to stellar cores with extremely slow rotation prior to collapse, hence BHs with low spins ($\chi \sim 0.01$, e.g., [232–234]). Though particular phases of mass transfer early in the evolution of the primary star can potentially lead to significant spin at birth [235], modeling of this evolutionary pathway finds that it does not lead to systems that can merge as BBHs [235,236]. Binary interactions following the formation of the first-born BH, such as mass transfer from the companion star to the BH, are also inefficient at spinning up BHs significantly [237–241]. On the other hand, the naked helium star precursor of the second-born BH in an isolated binary system can potentially be spun up through tidal interactions with the already-born BH if the system is in a tight enough orbital configuration [234,242–246]. Hence, there is a range of expectations for the spin magnitudes of both the primary and secondary components of BBHs formed in isolation.

As shown in Fig. 6, we find GW190412 to be consistent with a moderately spinning primary, though the spin of the secondary is broad and unconstrained. This interpretation is particularly apparent with the EOBNR PHM approximant, which finds the primary BH to have a dimensionless spin above 0.3 with 90% credibility regardless of the spin of the secondary. As this measurement indicates that at least *some* first-born BHs can be spinning significantly at birth, it will be useful for constraining theoretical uncertainties of BH natal spins if GW190412 is indeed the product of isolated binary evolution. An alternative interpretation has been put forth using a strong prior motivated by the assumption that primary BH is nonspinning, which leads to an inferred high spin of the secondary BH [247]. However, investigation into various prior assumptions for GW190412's spins has found the nonspinning primary interpretation to be disfavored by the data [248].

Hierarchical mergers in dynamical environments are a predicted pathway for forming BHs with significant spin. Even for BHs with negligible spin at birth, the product of an equal-mass hierarchical merger will attain a dimensionless spin of $\sim 0.7$ due to the angular momentum in the premerger orbit, e.g., [249]. As the primary spin magnitude of GW190412 is constrained to be $\lesssim 0.6$, the first-generation merger that formed the primary BH would need to have spins antialigned with the orbital angular momentum or be the merger of an unequal-mass system itself if the primary is indeed a second-generation merger product [170,208, 209,250]. However, hierarchical mergers are predicted to be relatively rare in dynamical environments compared to first-generation mergers. For example, in globular clusters hierarchical mergers make up only $\mathcal{O}(10\%)$ of detectable mergers even in the most optimistic cases, e.g., [209,210].

GW190412 also exhibits marginal signs of orbital precession. As shown in Fig. 5, we recover a posterior for $\chi_\mathrm{p}$ that deviates from the prior, hinting at GW190412 having some degree of in-plane spin. Though isolated binaries are predicted to lead to BBHs with preferentially aligned spins due to prior binary interactions such as mass transfer and tides, supernova natal kicks can lead to some misalignment of spins relative to the orbital plane (for low-mass BHs with significant kicks, 90% of systems form with tilts of $\lesssim 30°$ [251]). Furthermore, these kicks are predicted to be damped as the remnant mass increases due to mass fallback [252,253]. If formed from an isolated binary, GW190412 may be able to place constraints on the kicks necessary to explain the orbital precession present in the signal.

In summary, though the mass ratio of GW190412 is the most extreme of any BBH observed to date, it is consistent with the mass ratios predicted from a number of proposed BBH formation channels. Many astrophysical channels predict that near-equal-mass BBH mergers are more common than mergers with significantly asymmetric component masses. However, as the observational sample of BBHs grows, it is not unexpected that we would observe a system such as GW190412 that occupies a less probable region of intrinsic parameter space. In isolated binary evolution, the spins for the first-born BH in merging BBH systems are predicted to be small. This may indicate that the natal spins of BHs formed in isolation need to be revised, or that an alternative formation scenario is responsible for forming GW190412. Future detections of BBHs will enable tighter constraints on the rate of GW190412-like systems.

## VIII. CONCLUSIONS

Every observing run in the advanced GW detector era has delivered new science. After the first observations of BBHs in the first observing run, and the continued observation of BBHs as well as the multimessenger observation of a binary neutron star in the second observing run [6,254], O3 has been digging deeper into the populations of compact binary mergers. The observation of a likely second neutron-star binary in O3 has already been published [8], and here we presented another GW observation with previously unobserved features.

GW190412 was a highly significant event, with a combined SNR of 19 across all three GW detectors. It is the first binary observed that consists of two BHs of significantly asymmetric component masses. With 99% probability, the primary BH has more than twice the mass of its lighter companion. The measurement of asymmetric masses is also robust even when the properties of GW190412 are inferred using a population-based prior. This observation indicates that the astrophysical BBH population includes unequal-mass systems.

GW190412 is also a rich source from a more fundamental point of view. GR dictates that gravitational radiation from compact binaries is dominated by a quadrupolar structure, but it also contains weaker contributions from subdominant multipoles. Here we provided conclusive evidence that at least the second most important multipole—the $(\ell, |m|) = (3, 3)$ multipole—makes a significant, measurable contribution to the observed data. As a result, the orientation of the binary is more accurately determined and tighter bounds are obtained on relevant intrinsic source parameters such as the mass ratio and spin of the system.

The asymmetric mass ratio of GW190412 allows the primary spin to have a more measurable effect on the signal. We find the primary spin magnitude to be $0.44^{+0.16}_{-0.22}$, which is the strongest constraint on the individual spin magnitude of a BH using GWs so far. Though we only find marginal statistical hints of precession in the data, the results presented here illustrate that we confidently disfavor strong precession (as would be characterized by a large in-plane spin parameter).

GW190412 is a BBH that occupies a previously unobserved region of parameter space. As we continue to increase the sensitivity of our detectors and the time spent observing, we will gain a more complete picture of the BBH population. Future observations of similar types of binaries, or even more extreme mass ratios, will sharpen our understanding of their abundance and might help constrain formation mechanisms for such systems. GW190412 also shows that numerical and analytical advances in modeling coalescing binaries in previously unexplored regimes remains crucial for the analysis of current and future GW data. The most recent and most complete signal models robustly identified GW190412's source properties and showed consistency with GR. Systematic differences are visible and will become more important when we observe stronger signals, pointing to the necessity for future work in this area.

LIGO and Virgo data containing GW190412, and samples from a subset of the posterior probability distributions of the source parameters [255] (curated using the PESummary tool [256]), are available from the Gravitational Wave Open Science Center [108,257].

## ACKNOWLEDGMENTS

The authors gratefully acknowledge the support of the United States National Science Foundation (NSF) for the construction and operation of the LIGO Laboratory and Advanced LIGO as well as the Science and Technology Facilities Council (STFC) of the United Kingdom, the Max-Planck-Society (MPS), and the State of Niedersachsen/Germany for support of the construction of Advanced LIGO and construction and operation of the GEO600 detector. Additional support for Advanced LIGO was provided by the Australian Research Council. The authors gratefully acknowledge the Italian Istituto Nazionale di Fisica Nucleare (INFN), the French Centre National de la Recherche Scientifique (CNRS) and the Netherlands Organization for Scientific Research, for the construction and operation of the Virgo detector and the creation and support of the EGO consortium. The authors also gratefully acknowledge research support from these agencies as well as by the Council of Scientific and Industrial Research of India, the Department of Science and Technology, India, the Science & Engineering Research Board (SERB), India, the Ministry of Human Resource Development, India, the Spanish Agencia Estatal de Investigación, the Vicepresidència i Conselleria d'Innovació, Recerca i Turisme and the Conselleria d'Educació i Universitat del Govern de les Illes Balears, the Conselleria d'Innovació, Universitats, Ciència i Societat Digital de la Generalitat Valenciana and the CERCA Programme Generalitat de Catalunya, Spain, the National Science Centre of Poland, the Swiss National Science Foundation (SNSF), the Russian Foundation for Basic Research, the Russian Science Foundation, the European Commission, the European Regional Development Funds (ERDF), the Royal Society, the Scottish Funding Council, the Scottish Universities Physics Alliance, the Hungarian Scientific Research Fund (OTKA), the French Lyon Institute of Origins (LIO), the Belgian Fonds de la Recherche Scientifique (FRS-FNRS), Actions de Recherche Concertées (ARC) and Fonds Wetenschappelijk Onderzoek–Vlaanderen (FWO), Belgium, the Paris Île-de-France Region, the National Research, Development and Innovation Office Hungary (NKFIH), the National Research Foundation of Korea, Industry Canada and the Province of Ontario through the Ministry of Economic Development and Innovation, the Natural Science and Engineering Research Council Canada, the Canadian Institute for Advanced Research, the Brazilian Ministry of Science, Technology, Innovations, and Communications, the International Center for Theoretical Physics South American Institute for Fundamental Research (ICTP-SAIFR), the Research Grants Council of Hong Kong, the National Natural Science Foundation of China (NSFC), the Leverhulme Trust, the Research Corporation, the Ministry of Science and Technology (MOST), Taiwan and the Kavli Foundation. The authors gratefully acknowledge the support of the NSF, STFC, INFN and CNRS for provision of computational resources. We also thank the anonymous referees for their valuable comments on this paper.

R. Abbott,[1] T. D. Abbott,[2] S. Abraham,[3] F. Acernese,[4,5] K. Ackley,[6] C. Adams,[7] R. X. Adhikari,[1] V. B. Adya,[8] C. Affeldt,[9,10] M. Agathos,[11,12] K. Agatsuma,[13] N. Aggarwal,[14] O. D. Aguiar,[15] A. Aich,[16] L. Aiello,[17,18] A. Ain,[3] P. Ajith,[19] S. Akcay,[11]

G. Allen,[20] A. Allocca,[21] P. A. Altin,[8] A. Amato,[22] S. Anand,[1] A. Ananyeva,[1] S. B. Anderson,[1] W. G. Anderson,[23]
S. V. Angelova,[24] S. Ansoldi,[25,26] S. Antier,[27] S. Appert,[1] K. Arai,[1] M. C. Araya,[1] J. S. Areeda,[28] M. Arène,[27] N. Arnaud,[29,30]
S. M. Aronson,[31] K. G. Arun,[32] Y. Asali,[33] S. Ascenzi,[17,34] G. Ashton,[6] S. M. Aston,[7] P. Astone,[35] F. Aubin,[36] P. Aufmuth,[10]
K. AultONeal,[37] C. Austin,[2] V. Avendano,[38] S. Babak,[27] P. Bacon,[27] F. Badaracco,[17,18] M. K. M. Bader,[39] S. Bae,[40]
A. M. Baer,[41] J. Baird,[27] F. Baldaccini,[42,43] G. Ballardin,[30] S. W. Ballmer,[44] A. Bals,[37] A. Balsamo,[41] G. Baltus,[45]
S. Banagiri,[46] D. Bankar,[3] R. S. Bankar,[3] J. C. Barayoga,[1] C. Barbieri,[47,48] B. C. Barish,[1] D. Barker,[49] K. Barkett,[50]
P. Barneo,[51] F. Barone,[52,5] B. Barr,[53] L. Barsotti,[54] M. Barsuglia,[27] D. Barta,[55] J. Bartlett,[49] I. Bartos,[31] R. Bassiri,[56]
A. Basti,[57,21] M. Bawaj,[58,43] J. C. Bayley,[53] M. Bazzan,[59,60] B. Bécsy,[61] M. Bejger,[62] I. Belahcene,[29] A. S. Bell,[53]
D. Beniwal,[63] M. G. Benjamin,[37] R. Benkel,[64] J. D. Bentley,[13] F. Bergamin,[9] B. K. Berger,[56] G. Bergmann,[9,10] S. Bernuzzi,[11]
C. P. L. Berry,[14] D. Bersanetti,[65] A. Bertolini,[39] J. Betzwieser,[7] R. Bhandare,[66] A. V. Bhandari,[3] J. Bidler,[28] E. Biggs,[23]
I. A. Bilenko,[67] G. Billingsley,[1] R. Birney,[68] O. Birnholtz,[69,70] S. Biscans,[1,54] M. Bischi,[71,72] S. Biscoveanu,[54] A. Bisht,[10]
G. Bissenbayeva,[16] M. Bitossi,[30,21] M. A. Bizouard,[73] J. K. Blackburn,[1] J. Blackman,[50] C. D. Blair,[7] D. G. Blair,[74]
R. M. Blair,[49] F. Bobba,[75,76] N. Bode,[9,10] M. Boer,[73] Y. Boetzel,[77] G. Bogaert,[73] F. Bondu,[78] E. Bonilla,[56] R. Bonnand,[36]
P. Booker,[9,10] B. A. Boom,[39] R. Bork,[1] V. Boschi,[21] S. Bose,[3] V. Bossilkov,[74] J. Bosveld,[74] Y. Bouffanais,[59,60] A. Bozzi,[30]
C. Bradaschia,[21] P. R. Brady,[23] A. Bramley,[7] M. Branchesi,[17,18] J. E. Brau,[79] M. Breschi,[11] T. Briant,[80] J. H. Briggs,[53]
F. Brighenti,[71,72] A. Brillet,[73] M. Brinkmann,[9,10] R. Brito,[81,35,64] P. Brockill,[23] A. F. Brooks,[1] J. Brooks,[30] D. D. Brown,[63]
S. Brunett,[1] G. Bruno,[82] R. Bruntz,[41] A. Buikema,[54] T. Bulik,[83] H. J. Bulten,[84,39] A. Buonanno,[64,85] D. Buskulic,[36]
R. L. Byer,[56] M. Cabero,[9,10] L. Cadonati,[86] G. Cagnoli,[87] C. Cahillane,[1] J. Calderón Bustillo,[6] J. D. Callaghan,[53]
T. A. Callister,[1] E. Calloni,[88,5] J. B. Camp,[89] M. Canepa,[90,65] K. C. Cannon,[91] H. Cao,[63] J. Cao,[92] G. Carapella,[75,76]
F. Carbognani,[30] S. Caride,[93] M. F. Carney,[14] G. Carullo,[57,21] J. Casanueva Diaz,[21] C. Casentini,[94,34] J. Castañeda,[51]
S. Caudill,[39] M. Cavaglià,[95] F. Cavalier,[29] R. Cavalieri,[30] G. Cella,[21] P. Cerdá-Durán,[96] E. Cesarini,[97,34] O. Chaibi,[73]
K. Chakravarti,[3] C. Chan,[91] M. Chan,[53] S. Chao,[98] P. Charlton,[99] E. A. Chase,[14] E. Chassande-Mottin,[27] D. Chatterjee,[23]
M. Chaturvedi,[66] K. Chatziioannou,[100,101] H. Y. Chen,[102] X. Chen,[74] Y. Chen,[50] H.-P. Cheng,[31] C. K. Cheong,[103]
H. Y. Chia,[31] F. Chiadini,[104,76] R. Chierici,[105] A. Chincarini,[65] A. Chiummo,[30] G. Cho,[106] H. S. Cho,[107] M. Cho,[85]
N. Christensen,[73] Q. Chu,[74] S. Chua,[80] K. W. Chung,[103] S. Chung,[74] G. Ciani,[59,60] P. Ciecielag,[62] M. Cieślar,[62]
A. A. Ciobanu,[63] R. Ciolfi,[108,60] F. Cipriano,[73] A. Cirone,[90,65] F. Clara,[49] J. A. Clark,[86] P. Clearwater,[109] S. Clesse,[82]
F. Cleva,[73] E. Coccia,[17,18] P.-F. Cohadon,[80] D. Cohen,[29] M. Colleoni,[110] C. G. Collette,[111] C. Collins,[13] M. Colpi,[47,48]
M. Constancio Jr.,[15] L. Conti,[60] S. J. Cooper,[13] P. Corban,[7] T. R. Corbitt,[2] I. Cordero-Carrión,[112] S. Corezzi,[42,43]
K. R. Corley,[33] N. Cornish,[61] D. Corre,[29] A. Corsi,[93] S. Cortese,[30] C. A. Costa,[15] R. Cotesta,[64] M. W. Coughlin,[1]
S. B. Coughlin,[113,14] J.-P. Coulon,[73] S. T. Countryman,[33] P. Couvares,[1] P. B. Covas,[110] D. M. Coward,[74] M. J. Cowart,[7]
D. C. Coyne,[1] R. Coyne,[114] J. D. E. Creighton,[23] T. D. Creighton,[16] J. Cripe,[2] M. Croquette,[80] S. G. Crowder,[115]
J.-R. Cudell,[45] T. J. Cullen,[2] A. Cumming,[53] R. Cummings,[53] L. Cunningham,[53] E. Cuoco,[30] M. Curylo,[83] T. Dal Canton,[64]
G. Dálya,[116] A. Dana,[56] L. M. Daneshgaran-Bajastani,[117] B. D'Angelo,[90,65] S. L. Danilishin,[9,10] S. D'Antonio,[34]
K. Danzmann,[10,9] C. Darsow-Fromm,[118] A. Dasgupta,[119] L. E. H. Datrier,[53] V. Dattilo,[30] I. Dave,[66] M. Davier,[29]
G. S. Davies,[120] D. Davis,[44] E. J. Daw,[121] D. DeBra,[56] M. Deenadayalan,[3] J. Degallaix,[22] M. De Laurentis,[88,5] S. Deléglise,[80]
M. Delfavero,[69] N. De Lillo,[53] W. Del Pozzo,[57,21] L. M. DeMarchi,[14] V. D'Emilio,[113] N. Demos,[54] T. Dent,[120]
R. De Pietri,[122,123] R. De Rosa,[88,5] C. De Rossi,[30] R. DeSalvo,[124] O. de Varona,[9,10] S. Dhurandhar,[3] M. C. Díaz,[16]
M. Diaz-Ortiz Jr.,[31] T. Dietrich,[39] L. Di Fiore,[5] C. Di Fronzo,[13] C. Di Giorgio,[75,76] F. Di Giovanni,[96] M. Di Giovanni,[125,126]
T. Di Girolamo,[88,5] A. Di Lieto,[57,21] B. Ding,[111] S. Di Pace,[81,35] I. Di Palma,[81,35] F. Di Renzo,[57,21] A. K. Divakarla,[31]
A. Dmitriev,[13] Z. Doctor,[102] F. Donovan,[54] K. L. Dooley,[113] S. Doravari,[3] I. Dorrington,[113] T. P. Downes,[23] M. Drago,[17,18]
J. C. Driggers,[49] Z. Du,[92] J.-G. Ducoin,[29] P. Dupej,[53] O. Durante,[75,76] D. D'Urso,[127,128] S. E. Dwyer,[49] P. J. Easter,[6]
G. Eddolls,[53] B. Edelman,[79] T. B. Edo,[121] O. Edy,[129] A. Effler,[7] P. Ehrens,[1] J. Eichholz,[8] S. S. Eikenberry,[31] M. Eisenmann,[36]
R. A. Eisenstein,[54] A. Ejlli,[113] L. Errico,[88,5] R. C. Essick,[102] H. Estelles,[110] D. Estevez,[36] Z. B. Etienne,[130] T. Etzel,[1]
M. Evans,[54] T. M. Evans,[7] B. E. Ewing,[131] V. Fafone,[94,34,17] S. Fairhurst,[113] X. Fan,[92] S. Farinon,[65] B. Farr,[79]
W. M. Farr,[100,101] E. J. Fauchon-Jones,[113] M. Favata,[38] M. Fays,[121] M. Fazio,[132] J. Feicht,[1] M. M. Fejer,[56] F. Feng,[27]
E. Fenyvesi,[55,133] D. L. Ferguson,[86] A. Fernandez-Galiana,[54] I. Ferrante,[57,21] E. C. Ferreira,[15] T. A. Ferreira,[15]
F. Fidecaro,[57,21] I. Fiori,[30] D. Fiorucci,[17,18] M. Fishbach,[102] R. P. Fisher,[41] R. Fittipaldi,[134,76] M. Fitz-Axen,[46]
V. Fiumara,[135,76] R. Flaminio,[36,136] E. Floden,[46] E. Flynn,[28] H. Fong,[91] J. A. Font,[96,137] P. W. F. Forsyth,[8] J.-D. Fournier,[73]
S. Frasca,[81,35] F. Frasconi,[21] Z. Frei,[116] A. Freise,[13] R. Frey,[79] V. Frey,[29] P. Fritschel,[54] V. V. Frolov,[7] G. Fronzè,[138] P. Fulda,[31]

M. Fyffe,[7] H. A. Gabbard,[53] B. U. Gadre,[64] S. M. Gaebel,[13] J. R. Gair,[64] S. Galaudage,[6] D. Ganapathy,[54] A. Ganguly,[19] S. G. Gaonkar,[3] C. García-Quirós,[110] F. Garufi,[88,5] B. Gateley,[49] S. Gaudio,[37] V. Gayathri,[139] G. Gemme,[65] E. Genin,[30] A. Gennai,[21] D. George,[20] J. George,[66] L. Gergely,[140] S. Ghonge,[86] Abhirup Ghosh,[64] Archisman Ghosh,[141,142,143,39] S. Ghosh,[23] B. Giacomazzo,[125,126] J. A. Giaime,[2,7] K. D. Giardina,[7] D. R. Gibson,[68] C. Gier,[24] K. Gill,[33] J. Glanzer,[2] J. Gniesmer,[118] P. Godwin,[131] E. Goetz,[2,95] R. Goetz,[31] N. Gohlke,[9,10] B. Goncharov,[6] G. González,[2] A. Gopakumar,[144] S. E. Gossan,[1] M. Gosselin,[30,57,21] R. Gouaty,[36] B. Grace,[8] A. Grado,[145,5] M. Granata,[22] A. Grant,[53] S. Gras,[54] P. Grassia,[1] C. Gray,[49] R. Gray,[53] G. Greco,[71,72] A. C. Green,[31] R. Green,[113] E. M. Gretarsson,[37] H. L. Griggs,[86] G. Grignani,[42,43] A. Grimaldi,[125,126] S. J. Grimm,[17,18] H. Grote,[113] S. Grunewald,[64] P. Gruning,[29] G. M. Guidi,[71,72] A. R. Guimaraes,[2] G. Guixé,[51] H. K. Gulati,[119] Y. Guo,[39] A. Gupta,[131] Anchal Gupta,[1] P. Gupta,[39] E. K. Gustafson,[1] R. Gustafson,[146] L. Haegel,[110] O. Halim,[18,17] E. D. Hall,[54] E. Z. Hamilton,[113] G. Hammond,[53] M. Haney,[77] M. M. Hanke,[9,10] J. Hanks,[49] C. Hanna,[131] M. D. Hannam,[113] O. A. Hannuksela,[103] T. J. Hansen,[37] J. Hanson,[7] T. Harder,[73] T. Hardwick,[2] K. Haris,[19] J. Harms,[17,18] G. M. Harry,[147] I. W. Harry,[129] R. K. Hasskew,[7] C.-J. Haster,[54] K. Haughian,[53] F. J. Hayes,[53] J. Healy,[69] A. Heidmann,[80] M. C. Heintze,[7] J. Heinze,[9,10] H. Heitmann,[73] F. Hellman,[148] P. Hello,[29] G. Hemming,[30] M. Hendry,[53] I. S. Heng,[53] E. Hennes,[39] J. Hennig,[9,10] M. Heurs,[9,10] S. Hild,[149,53] T. Hinderer,[143,39,141] S. Y. Hoback,[28,147] S. Hochheim,[9,10] E. Hofgard,[56] D. Hofman,[22] A. M. Holgado,[20] N. A. Holland,[8] K. Holt,[7] D. E. Holz,[102] P. Hopkins,[113] C. Horst,[23] J. Hough,[53] E. J. Howell,[74] C. G. Hoy,[113] Y. Huang,[54] M. T. Hübner,[6] E. A. Huerta,[20] D. Huet,[29] B. Hughey,[37] V. Hui,[36] S. Husa,[110] S. H. Huttner,[53] R. Huxford,[131] T. Huynh-Dinh,[7] B. Idzkowski,[83] A. Iess,[94,34] H. Inchauspe,[31] C. Ingram,[63] G. Intini,[81,35] J.-M. Isac,[80] M. Isi,[54] B. R. Iyer,[19] T. Jacqmin,[80] S. J. Jadhav,[150] S. P. Jadhav,[3] A. L. James,[113] K. Jani,[86] N. N. Janthalur,[150] P. Jaranowski,[151] D. Jariwala,[31] R. Jaume,[110] A. C. Jenkins,[152] J. Jiang,[31] G. R. Johns,[41] N. K. Johnson-McDaniel,[12] A. W. Jones,[13] D. I. Jones,[153] J. D. Jones,[49] P. Jones,[13] R. Jones,[53] R. J. G. Jonker,[39] L. Ju,[74] J. Junker,[9,10] C. V. Kalaghatgi,[113] V. Kalogera,[14] B. Kamai,[1] S. Kandhasamy,[3] G. Kang,[40] J. B. Kanner,[1] S. J. Kapadia,[19] S. Karki,[79] R. Kashyap,[19] M. Kasprzack,[1] W. Kastaun,[9,10] S. Katsanevas,[30] E. Katsavounidis,[54] W. Katzman,[7] S. Kaufer,[10] K. Kawabe,[49] F. Kéfélian,[73] D. Keitel,[129] A. Keivani,[33] R. Kennedy,[121] J. S. Key,[154] S. Khadka,[56] F. Y. Khalili,[67] I. Khan,[17,34] S. Khan,[9,10] Z. A. Khan,[92] E. A. Khazanov,[155] N. Khetan,[17,18] M. Khursheed,[66] N. Kijbunchoo,[8] Chunglee Kim,[156] G. J. Kim,[86] J. C. Kim,[157] K. Kim,[103] W. Kim,[63] W. S. Kim,[158] Y.-M. Kim,[159] C. Kimball,[14] P. J. King,[49] M. Kinley-Hanlon,[53] R. Kirchhoff,[9,10] J. S. Kissel,[49] L. Kleybolte,[118] S. Klimenko,[31] T. D. Knowles,[130] E. Knyazev,[54] P. Koch,[9,10] S. M. Koehlenbeck,[9,10] G. Koekoek,[39,149] S. Koley,[39] V. Kondrashov,[1] A. Kontos,[160] N. Koper,[9,10] M. Korobko,[118] W. Z. Korth,[1] M. Kovalam,[74] D. B. Kozak,[1] V. Kringel,[9,10] N. V. Krishnendu,[9,10] A. Królak,[161,162] N. Krupinski,[23] G. Kuehn,[9,10] A. Kumar,[150] P. Kumar,[163] Rahul Kumar,[49] Rakesh Kumar,[119] S. Kumar,[19] L. Kuo,[98] A. Kutynia,[161] B. D. Lackey,[64] D. Laghi,[57,21] E. Lalande,[164] T. L. Lam,[103] A. Lamberts,[73,165] M. Landry,[49] B. B. Lane,[54] R. N. Lang,[166] J. Lange,[69] B. Lantz,[56] R. K. Lanza,[54] I. La Rosa,[36] A. Lartaux-Vollard,[29] P. D. Lasky,[6] M. Laxen,[7] A. Lazzarini,[1] C. Lazzaro,[60] P. Leaci,[81,35] S. Leavey,[9,10] Y. K. Lecoeuche,[49] C. H. Lee,[107] H. M. Lee,[167] H. W. Lee,[157] J. Lee,[106] K. Lee,[56] J. Lehmann,[9,10] N. Leroy,[29] N. Letendre,[36] Y. Levin,[6] A. K. Y. Li,[103] J. Li,[92] K. Li,[103] T. G. F. Li,[103] X. Li,[50] F. Linde,[168,39] S. D. Linker,[117] J. N. Linley,[53] T. B. Littenberg,[169] J. Liu,[9,10] X. Liu,[23] M. Llorens-Monteagudo,[96] R. K. L. Lo,[1] A. Lockwood,[170] L. T. London,[54] A. Longo,[171,172] M. Lorenzini,[17,18] V. Loriette,[173] M. Lormand,[7] G. Losurdo,[21] J. D. Lough,[9,10] C. O. Lousto,[69] G. Lovelace,[28] H. Lück,[10,9] D. Lumaca,[94,34] A. P. Lundgren,[129] Y. Ma,[50] R. Macas,[113] S. Macfoy,[24] M. MacInnis,[54] D. M. Macleod,[113] I. A. O. MacMillan,[147] A. Macquet,[73] I. Magaña Hernandez,[23] F. Magaña-Sandoval,[31] R. M. Magee,[131] E. Majorana,[35] I. Maksimovic,[173] A. Malik,[66] N. Man,[73] V. Mandic,[46] V. Mangano,[53,81,35] G. L. Mansell,[49,54] M. Manske,[23] M. Mantovani,[30] M. Mapelli,[59,60] F. Marchesoni,[58,43,174] F. Marion,[36] S. Márka,[33] Z. Márka,[33] C. Markakis,[12] A. S. Markosyan,[56] A. Markowitz,[1] E. Maros,[1] A. Marquina,[112] S. Marsat,[27] F. Martelli,[71,72] I. W. Martin,[53] R. M. Martin,[38] V. Martinez,[87] D. V. Martynov,[13] H. Masalehdan,[118] K. Mason,[54] E. Massera,[121] A. Masserot,[36] T. J. Massinger,[54] M. Masso-Reid,[53] S. Mastrogiovanni,[27] A. Matas,[64] F. Matichard,[1,54] N. Mavalvala,[54] E. Maynard,[2] J. J. McCann,[74] R. McCarthy,[49] D. E. McClelland,[8] S. McCormick,[7] L. McCuller,[54] S. C. McGuire,[175] C. McIsaac,[129] J. McIver,[1] D. J. McManus,[8] T. McRae,[8] S. T. McWilliams,[130] D. Meacher,[23] G. D. Meadors,[6] M. Mehmet,[9,10] A. K. Mehta,[19] E. Mejuto Villa,[124,76] A. Melatos,[109] G. Mendell,[49] R. A. Mercer,[23] L. Mereni,[22] K. Merfeld,[79] E. L. Merilh,[49] J. D. Merritt,[79] M. Merzougui,[73] S. Meshkov,[1] C. Messenger,[53] C. Messick,[176] R. Metzdorff,[80] P. M. Meyers,[109] F. Meylahn,[9,10] A. Mhaske,[3] A. Miani,[125,126] H. Miao,[13] I. Michaloliakos,[31] C. Michel,[22] H. Middleton,[109] L. Milano,[88,5] A. L. Miller,[31,81,35] S. Miller,[1] M. Millhouse,[109] J. C. Mills,[113] E. Milotti,[177,26] M. C. Milovich-Goff,[117] O. Minazzoli,[73,178] Y. Minenkov,[34] A. Mishkin,[31] C. Mishra,[179] T. Mistry,[121] S. Mitra,[3] V. P. Mitrofanov,[67] G. Mitselmakher,[31] R. Mittleman,[54] G. Mo,[54]

K. Mogushi,[95] S. R. P. Mohapatra,[54] S. R. Mohite,[23] M. Molina-Ruiz,[148] M. Mondin,[117] M. Montani,[71,72] C. J. Moore,[13] D. Moraru,[49] F. Morawski,[62] G. Moreno,[49] S. Morisaki,[91] B. Mours,[180] C. M. Mow-Lowry,[13] S. Mozzon,[129] F. Muciaccia,[81,35] Arunava Mukherjee,[53] D. Mukherjee,[131] S. Mukherjee,[16] Subroto Mukherjee,[119] N. Mukund,[9,10] A. Mullavey,[7] J. Munch,[63] E. A. Muñiz,[44] P. G. Murray,[53] A. Nagar,[97,138,181] I. Nardecchia,[94,34] L. Naticchioni,[81,35] R. K. Nayak,[182] B. F. Neil,[74] J. Neilson,[124,76] G. Nelemans,[183,39] T. J. N. Nelson,[7] M. Nery,[9,10] A. Neunzert,[146] K. Y. Ng,[54] S. Ng,[63] C. Nguyen,[27] P. Nguyen,[79] D. Nichols,[143,39] S. A. Nichols,[2] S. Nissanke,[143,39] F. Nocera,[30] M. Noh,[54] C. North,[113] D. Nothard,[184] L. K. Nuttall,[129] J. Oberling,[49] B. D. O'Brien,[31] G. Oganesyan,[17,18] G. H. Ogin,[185] J. J. Oh,[158] S. H. Oh,[158] F. Ohme,[9,10] H. Ohta,[91] M. A. Okada,[15] M. Oliver,[110] C. Olivetto,[30] P. Oppermann,[9,10] Richard J. Oram,[7] B. O'Reilly,[7] R. G. Ormiston,[46] L. F. Ortega,[31] R. O'Shaughnessy,[69] S. Ossokine,[64] C. Osthelder,[1] D. J. Ottaway,[63] H. Overmier,[7] B. J. Owen,[93] A. E. Pace,[131] G. Pagano,[57,21] M. A. Page,[74] G. Pagliaroli,[17,18] A. Pai,[139] S. A. Pai,[66] J. R. Palamos,[79] O. Palashov,[155] C. Palomba,[35] H. Pan,[98] P. K. Panda,[150] P. T. H. Pang,[39] C. Pankow,[14] F. Pannarale,[81,35] B. C. Pant,[66] F. Paoletti,[21] A. Paoli,[30] A. Parida,[3] W. Parker,[7,175] D. Pascucci,[53,39] A. Pasqualetti,[30] R. Passaquieti,[57,21] D. Passuello,[21] B. Patricelli,[57,21] E. Payne,[6] B. L. Pearlstone,[53] T. C. Pechsiri,[31] A. J. Pedersen,[44] M. Pedraza,[1] A. Pele,[7] S. Penn,[186] A. Perego,[125,126] C. J. Perez,[49] C. Périgois,[36] A. Perreca,[125,126] S. Perriès,[105] J. Petermann,[118] H. P. Pfeiffer,[64] M. Phelps,[9,10] K. S. Phukon,[3,168,39] O. J. Piccinni,[81,35] M. Pichot,[73] M. Piendibene,[57,21] F. Piergiovanni,[71,72] V. Pierro,[124,76] G. Pillant,[30] L. Pinard,[22] I. M. Pinto,[124,76,97] K. Piotrzkowski,[82] M. Pirello,[49] M. Pitkin,[187] W. Plastino,[171,172] R. Poggiani,[57,21] D. Y. T. Pong,[103] S. Ponrathnam,[3] P. Popolizio,[30] E. K. Porter,[27] J. Powell,[188] A. K. Prajapati,[119] K. Prasai,[56] R. Prasanna,[150] G. Pratten,[13] T. Prestegard,[23] M. Principe,[124,97,76] G. A. Prodi,[125,126] L. Prokhorov,[13] M. Punturo,[43] P. Puppo,[35] M. Pürrer,[64] H. Qi,[113] V. Quetschke,[16] P. J. Quinonez,[37] F. J. Raab,[49] G. Raaijmakers,[143,39] H. Radkins,[49] N. Radulesco,[73] P. Raffai,[116] H. Rafferty,[189] S. Raja,[66] C. Rajan,[66] B. Rajbhandari,[93] M. Rakhmanov,[16] K. E. Ramirez,[16] A. Ramos-Buades,[110] Javed Rana,[3] K. Rao,[14] P. Rapagnani,[81,35] V. Raymond,[113] M. Razzano,[57,21] J. Read,[28] T. Regimbau,[36] L. Rei,[65] S. Reid,[24] D. H. Reitze,[1,31] P. Rettegno,[138,190] F. Ricci,[81,35] C. J. Richardson,[37] J. W. Richardson,[1] P. M. Ricker,[20] G. Riemenschneider,[190,138] K. Riles,[146] M. Rizzo,[14] N. A. Robertson,[1,53] F. Robinet,[29] A. Rocchi,[34] R. D. Rodriguez-Soto,[37] L. Rolland,[36] J. G. Rollins,[1] V. J. Roma,[79] M. Romanelli,[78] R. Romano,[4,5] C. L. Romel,[49] I. M. Romero-Shaw,[6] J. H. Romie,[7] C. A. Rose,[23] D. Rose,[28] K. Rose,[184] D. Rosińska,[83] S. G. Rosofsky,[20] M. P. Ross,[170] S. Rowan,[53] S. J. Rowlinson,[13] P. K. Roy,[16] Santosh Roy,[3] Soumen Roy,[191] P. Ruggi,[30] G. Rutins,[68] K. Ryan,[49] S. Sachdev,[131] T. Sadecki,[49] M. Sakellariadou,[152] O. S. Salafia,[192,47,48] L. Salconi,[30] M. Saleem,[32] A. Samajdar,[39] E. J. Sanchez,[1] L. E. Sanchez,[1] N. Sanchis-Gual,[193] J. R. Sanders,[194] K. A. Santiago,[38] E. Santos,[73] N. Sarin,[6] B. Sassolas,[22] B. S. Sathyaprakash,[131,113] O. Sauter,[36] R. L. Savage,[49] V. Savant,[3] D. Sawant,[139] S. Sayah,[22] D. Schaetzl,[1] P. Schale,[79] M. Scheel,[50] J. Scheuer,[14] P. Schmidt,[13] R. Schnabel,[118] R. M. S. Schofield,[79] A. Schönbeck,[118] E. Schreiber,[9,10] B. W. Schulte,[9,10] B. F. Schutz,[113] O. Schwarm,[185] E. Schwartz,[7] J. Scott,[53] S. M. Scott,[8] E. Seidel,[20] D. Sellers,[7] A. S. Sengupta,[191] N. Sennett,[64] D. Sentenac,[30] V. Sequino,[65] A. Sergeev,[155] Y. Setyawati,[9,10] D. A. Shaddock,[8] T. Shaffer,[49] M. S. Shahriar,[14] S. Sharifi,[2] A. Sharma,[17,18] P. Sharma,[66] P. Shawhan,[85] H. Shen,[20] M. Shikauchi,[91] R. Shink,[164] D. H. Shoemaker,[54] D. M. Shoemaker,[86] K. Shukla,[148] S. ShyamSundar,[66] K. Siellez,[86] M. Sieniawska,[62] D. Sigg,[49] L. P. Singer,[89] D. Singh,[131] N. Singh,[83] A. Singha,[53] A. Singhal,[17,35] A. M. Sintes,[110] V. Sipala,[127,128] V. Skliris,[113] B. J. J. Slagmolen,[8] T. J. Slaven-Blair,[74] J. Smetana,[13] J. R. Smith,[28] R. J. E. Smith,[6] S. Somala,[195] E. J. Son,[158] S. Soni,[2] B. Sorazu,[53] V. Sordini,[105] F. Sorrentino,[65] T. Souradeep,[3] E. Sowell,[93] A. P. Spencer,[53] M. Spera,[59,60] A. K. Srivastava,[119] V. Srivastava,[44] K. Staats,[14] C. Stachie,[73] M. Standke,[9,10] D. A. Steer,[27] M. Steinke,[9,10] J. Steinlechner,[118,53] S. Steinlechner,[118] D. Steinmeyer,[9,10] S. Stevenson,[188] D. Stocks,[56] D. J. Stops,[13] M. Stover,[184] K. A. Strain,[53] G. Stratta,[196,72] A. Strunk,[49] R. Sturani,[197] A. L. Stuver,[198] S. Sudhagar,[3] V. Sudhir,[54] T. Z. Summerscales,[199] L. Sun,[1] S. Sunil,[119] A. Sur,[62] J. Suresh,[91] P. J. Sutton,[113] B. L. Swinkels,[39] M. J. Szczepańczyk,[31] M. Tacca,[39] S. C. Tait,[53] C. Talbot,[6] A. J. Tanasijczuk,[82] D. B. Tanner,[31] D. Tao,[1] M. Tápai,[140] A. Tapia,[28] E. N. Tapia San Martin,[39] J. D. Tasson,[200] R. Taylor,[1] R. Tenorio,[110] L. Terkowski,[118] M. P. Thirugnanasambandam,[3] M. Thomas,[7] P. Thomas,[49] J. E. Thompson,[113] S. R. Thondapu,[66] K. A. Thorne,[7] E. Thrane,[6] C. L. Tinsman,[6] T. R. Saravanan,[3] Shubhanshu Tiwari,[77,125,126] S. Tiwari,[144] V. Tiwari,[113] K. Toland,[53] M. Tonelli,[57,21] Z. Tornasi,[53] A. Torres-Forné,[64] C. I. Torrie,[1] I. Tosta e Melo,[127,128] D. Töyrä,[8] E. A. Trail,[2] F. Travasso,[58,43] G. Traylor,[7] M. C. Tringali,[83] A. Tripathee,[146] A. Trovato,[27] R. J. Trudeau,[1] K. W. Tsang,[39] M. Tse,[54] R. Tso,[50] L. Tsukada,[91] D. Tsuna,[91] T. Tsutsui,[91] M. Turconi,[73] A. S. Ubhi,[13] R. Udall,[86] K. Ueno,[91] D. Ugolini,[189] C. S. Unnikrishnan,[144] A. L. Urban,[2] S. A. Usman,[102] A. C. Utina,[53] H. Vahlbruch,[10] G. Vajente,[1] G. Valdes,[2] M. Valentini,[125,126] N. van Bakel,[39] M. van Beuzekom,[39] J. F. J. van den Brand,[84,149,39] C. Van Den Broeck,[39,201] D. C. Vander-Hyde,[44] L. van der Schaaf,[39]

J. V. Van Heijningen,[74] A. A. van Veggel,[53] M. Vardaro,[168,39] V. Varma,[50] S. Vass,[1] M. Vasúth,[55] A. Vecchio,[13] G. Vedovato,[60] J. Veitch,[53] P. J. Veitch,[63] K. Venkateswara,[170] G. Venugopalan,[1] D. Verkindt,[36] D. Veske,[33] F. Vetrano,[71,72] A. Viceré,[71,72] A. D. Viets,[202] S. Vinciguerra,[13] D. J. Vine,[68] J.-Y. Vinet,[73] S. Vitale,[54] Francisco Hernandez Vivanco,[6] T. Vo,[44] H. Vocca,[42,43] C. Vorvick,[49] S. P. Vyatchanin,[67] A. R. Wade,[8] L. E. Wade,[184] M. Wade,[184] R. Walet,[39] M. Walker,[28] G. S. Wallace,[24] L. Wallace,[1] S. Walsh,[23] J. Z. Wang,[146] S. Wang,[20] W. H. Wang,[16] R. L. Ward,[8] Z. A. Warden,[37] J. Warner,[49] M. Was,[36] J. Watchi,[111] B. Weaver,[49] L.-W. Wei,[9,10] M. Weinert,[9,10] A. J. Weinstein,[1] R. Weiss,[54] F. Wellmann,[9,10] L. Wen,[74] P. Weßels,[9,10] J. W. Westhouse,[37] K. Wette,[8] J. T. Whelan,[69] B. F. Whiting,[31] C. Whittle,[54] D. M. Wilken,[9,10] D. Williams,[53] J. L. Willis,[1] B. Willke,[10,9] W. Winkler,[9,10] C. C. Wipf,[1] H. Wittel,[9,10] G. Woan,[53] J. Woehler,[9,10] J. K. Wofford,[69] C. Wong,[103] J. L. Wright,[53] D. S. Wu,[9,10] D. M. Wysocki,[69] L. Xiao,[1] H. Yamamoto,[1] L. Yang,[132] Y. Yang,[31] Z. Yang,[46] M. J. Yap,[8] M. Yazback,[31] D. W. Yeeles,[113] Hang Yu,[54] Haocun Yu,[54] S. H. R. Yuen,[103] A. K. Zadrożny,[16] A. Zadrożny,[161] M. Zanolin,[37] T. Zelenova,[30] J.-P. Zendri,[60] M. Zevin,[14] J. Zhang,[74] L. Zhang,[1] T. Zhang,[53] C. Zhao,[74] G. Zhao,[111] M. Zhou,[14] Z. Zhou,[14] X. J. Zhu,[6] A. B. Zimmerman,[176] M. E. Zucker,[54,1] and J. Zweizig[1]

(LIGO Scientific Collaboration and Virgo Collaboration)

[1]*LIGO, California Institute of Technology, Pasadena, California 91125, USA*
[2]*Louisiana State University, Baton Rouge, Louisiana 70803, USA*
[3]*Inter-University Centre for Astronomy and Astrophysics, Pune 411007, India*
[4]*Dipartimento di Farmacia, Università di Salerno, I-84084 Fisciano, Salerno, Italy*
[5]*INFN, Sezione di Napoli, Complesso Universitario di Monte S. Angelo, I-80126 Napoli, Italy*
[6]*OzGrav, School of Physics & Astronomy, Monash University, Clayton 3800, Victoria, Australia*
[7]*LIGO Livingston Observatory, Livingston, Louisiana 70754, USA*
[8]*OzGrav, Australian National University, Canberra, Australian Capital Territory 0200, Australia*
[9]*Max Planck Institute for Gravitational Physics (Albert Einstein Institute), D-30167 Hannover, Germany*
[10]*Leibniz Universität Hannover, D-30167 Hannover, Germany*
[11]*Theoretisch-Physikalisches Institut, Friedrich-Schiller-Universität Jena, D-07743 Jena, Germany*
[12]*University of Cambridge, Cambridge CB2 1TN, United Kingdom*
[13]*University of Birmingham, Birmingham B15 2TT, United Kingdom*
[14]*Center for Interdisciplinary Exploration & Research in Astrophysics (CIERA), Northwestern University, Evanston, Illinois 60208, USA*
[15]*Instituto Nacional de Pesquisas Espaciais, 12227-010 São José dos Campos, São Paulo, Brazil*
[16]*The University of Texas Rio Grande Valley, Brownsville, Texas 78520, USA*
[17]*Gran Sasso Science Institute (GSSI), I-67100 L'Aquila, Italy*
[18]*INFN, Laboratori Nazionali del Gran Sasso, I-67100 Assergi, Italy*
[19]*International Centre for Theoretical Sciences, Tata Institute of Fundamental Research, Bengaluru 560089, India*
[20]*NCSA, University of Illinois at Urbana-Champaign, Urbana, Illinois 61801, USA*
[21]*INFN, Sezione di Pisa, I-56127 Pisa, Italy*
[22]*Laboratoire des Matériaux Avancés (LMA), IP2I–UMR 5822, CNRS, Université de Lyon, F-69622 Villeurbanne, France*
[23]*University of Wisconsin-Milwaukee, Milwaukee, Wisconsin 53201, USA*
[24]*SUPA, University of Strathclyde, Glasgow G1 1XQ, United Kingdom*
[25]*Dipartimento di Matematica e Informatica, Università di Udine, I-33100 Udine, Italy*
[26]*INFN, Sezione di Trieste, I-34127 Trieste, Italy*
[27]*APC, AstroParticule et Cosmologie, Université Paris Diderot, CNRS/IN2P3, CEA/Irfu, Observatoire de Paris, Sorbonne Paris Cité, F-75205 Paris Cedex 13, France*
[28]*California State University Fullerton, Fullerton, California 92831, USA*
[29]*LAL, Université Paris-Sud, CNRS/IN2P3, Université Paris-Saclay, F-91898 Orsay, France*
[30]*European Gravitational Observatory (EGO), I-56021 Cascina, Pisa, Italy*
[31]*University of Florida, Gainesville, Florida 32611, USA*
[32]*Chennai Mathematical Institute, Chennai 603103, India*
[33]*Columbia University, New York, New York 10027, USA*
[34]*INFN, Sezione di Roma Tor Vergata, I-00133 Roma, Italy*
[35]*INFN, Sezione di Roma, I-00185 Roma, Italy*
[36]*Laboratoire d'Annecy de Physique des Particules (LAPP), Université Grenoble Alpes, Université Savoie Mont Blanc, CNRS/IN2P3, F-74941 Annecy, France*
[37]*Embry-Riddle Aeronautical University, Prescott, Arizona 86301, USA*

[38]Montclair State University, Montclair, New Jersey 07043, USA
[39]Nikhef, Science Park 105, 1098 XG Amsterdam, The Netherlands
[40]Korea Institute of Science and Technology Information, Daejeon 34141, South Korea
[41]Christopher Newport University, Newport News, Virginia 23606, USA
[42]Università di Perugia, I-06123 Perugia, Italy
[43]INFN, Sezione di Perugia, I-06123 Perugia, Italy
[44]Syracuse University, Syracuse, New York 13244, USA
[45]Université de Liège, B-4000 Liège, Belgium
[46]University of Minnesota, Minneapolis, Minnesota 55455, USA
[47]Università degli Studi di Milano-Bicocca, I-20126 Milano, Italy
[48]INFN, Sezione di Milano-Bicocca, I-20126 Milano, Italy
[49]LIGO Hanford Observatory, Richland, Washington 99352, USA
[50]Caltech CaRT, Pasadena, California 91125, USA
[51]Departament de Física Quàntica i Astrofísica, Institut de Ciències del Cosmos (ICCUB), Universitat de Barcelona (IEEC-UB), E-08028 Barcelona, Spain
[52]Dipartimento di Medicina, Chirurgia e Odontoiatria "Scuola Medica Salernitana," Università di Salerno, I-84081 Baronissi, Salerno, Italy
[53]SUPA, University of Glasgow, Glasgow G12 8QQ, United Kingdom
[54]LIGO, Massachusetts Institute of Technology, Cambridge, Massachusetts 02139, USA
[55]Wigner RCP, RMKI, H-1121 Budapest, Konkoly Thege Miklós út 29-33, Hungary
[56]Stanford University, Stanford, California 94305, USA
[57]Università di Pisa, I-56127 Pisa, Italy
[58]Università di Camerino, Dipartimento di Fisica, I-62032 Camerino, Italy
[59]Università di Padova, Dipartimento di Fisica e Astronomia, I-35131 Padova, Italy
[60]INFN, Sezione di Padova, I-35131 Padova, Italy
[61]Montana State University, Bozeman, Montana 59717, USA
[62]Nicolaus Copernicus Astronomical Center, Polish Academy of Sciences, 00-716, Warsaw, Poland
[63]OzGrav, University of Adelaide, Adelaide, South Australia 5005, Australia
[64]Max Planck Institute for Gravitational Physics (Albert Einstein Institute), D-14476 Potsdam-Golm, Germany
[65]INFN, Sezione di Genova, I-16146 Genova, Italy
[66]RRCAT, Indore, Madhya Pradesh 452013, India
[67]Faculty of Physics, Lomonosov Moscow State University, Moscow 119991, Russia
[68]SUPA, University of the West of Scotland, Paisley PA1 2BE, United Kingdom
[69]Rochester Institute of Technology, Rochester, New York 14623, USA
[70]Bar-Ilan University, Ramat Gan 5290002, Israel
[71]Università degli Studi di Urbino "Carlo Bo," I-61029 Urbino, Italy
[72]INFN, Sezione di Firenze, I-50019 Sesto Fiorentino, Firenze, Italy
[73]Artemis, Université Côte d'Azur, Observatoire Côte d'Azur, CNRS, CS 34229, F-06304 Nice Cedex 4, France
[74]OzGrav, University of Western Australia, Crawley, Western Australia 6009, Australia
[75]Dipartimento di Fisica "E.R. Caianiello," Università di Salerno, I-84084 Fisciano, Salerno, Italy
[76]INFN, Sezione di Napoli, Gruppo Collegato di Salerno, Complesso Universitario di Monte S. Angelo, I-80126 Napoli, Italy
[77]Physik-Institut, University of Zurich, Winterthurerstrasse 190, 8057 Zurich, Switzerland
[78]Université Rennes, CNRS, Institut FOTON–UMR6082, F-3500 Rennes, France
[79]University of Oregon, Eugene, Oregon 97403, USA
[80]Laboratoire Kastler Brossel, Sorbonne Université, CNRS, ENS-Université PSL, Collège de France, F-75005 Paris, France
[81]Università di Roma "La Sapienza," I-00185 Roma, Italy
[82]Université catholique de Louvain, B-1348 Louvain-la-Neuve, Belgium
[83]Astronomical Observatory Warsaw University, 00-478 Warsaw, Poland
[84]VU University Amsterdam, 1081 HV Amsterdam, The Netherlands
[85]University of Maryland, College Park, Maryland 20742, USA
[86]School of Physics, Georgia Institute of Technology, Atlanta, Georgia 30332, USA
[87]Université de Lyon, Université Claude Bernard Lyon 1, CNRS, Institut Lumière Matière, F-69622 Villeurbanne, France
[88]Università di Napoli "Federico II," Complesso Universitario di Monte S. Angelo, I-80126 Napoli, Italy
[89]NASA Goddard Space Flight Center, Greenbelt, Maryland 20771, USA
[90]Dipartimento di Fisica, Università degli Studi di Genova, I-16146 Genova, Italy

[91]*RESCEU, University of Tokyo, Tokyo, 113-0033, Japan*
[92]*Tsinghua University, Beijing 100084, China*
[93]*Texas Tech University, Lubbock, Texas 79409, USA*
[94]*Università di Roma Tor Vergata, I-00133 Roma, Italy*
[95]*Missouri University of Science and Technology, Rolla, Missouri 65409, USA*
[96]*Departamento de Astronomía y Astrofísica, Universitat de València, E-46100 Burjassot, València, Spain*
[97]*Museo Storico della Fisica e Centro Studi e Ricerche "Enrico Fermi," I-00184 Roma, Italy*
[98]*National Tsing Hua University, Hsinchu City, 30013 Taiwan, Republic of China*
[99]*Charles Sturt University, Wagga Wagga, New South Wales 2678, Australia*
[100]*Physics and Astronomy Department, Stony Brook University, Stony Brook, New York 11794, USA*
[101]*Center for Computational Astrophysics, Flatiron Institute, 162 5th Avenue, New York, New York 10010, USA*
[102]*University of Chicago, Chicago, Illinois 60637, USA*
[103]*The Chinese University of Hong Kong, Shatin, New Territories, Hong Kong*
[104]*Dipartimento di Ingegneria Industriale (DIIN), Università di Salerno, I-84084 Fisciano, Salerno, Italy*
[105]*Institut de Physique des 2 Infinis de Lyon (IP2I)–UMR 5822, Université de Lyon, Université Claude Bernard, CNRS, F-69622 Villeurbanne, France*
[106]*Seoul National University, Seoul 08826, South Korea*
[107]*Pusan National University, Busan 46241, South Korea*
[108]*INAF, Osservatorio Astronomico di Padova, I-35122 Padova, Italy*
[109]*OzGrav, University of Melbourne, Parkville, Victoria 3010, Australia*
[110]*Universitat de les Illes Balears, IAC3–IEEC, E-07122 Palma de Mallorca, Spain*
[111]*Université Libre de Bruxelles, Brussels 1050, Belgium*
[112]*Departamento de Matemáticas, Universitat de València, E-46100 Burjassot, València, Spain*
[113]*Cardiff University, Cardiff CF24 3AA, United Kingdom*
[114]*University of Rhode Island, Kingston, Rhode Island 02881, USA*
[115]*Bellevue College, Bellevue, Washington 98007, USA*
[116]*MTA-ELTE Astrophysics Research Group, Institute of Physics, Eötvös University, Budapest 1117, Hungary*
[117]*California State University, Los Angeles, 5151 State University Drive, Los Angeles, California 90032, USA*
[118]*Universität Hamburg, D-22761 Hamburg, Germany*
[119]*Institute for Plasma Research, Bhat, Gandhinagar 382428, India*
[120]*IGFAE, Campus Sur, Universidade de Santiago de Compostela, 15782, Spain*
[121]*The University of Sheffield, Sheffield S10 2TN, United Kingdom*
[122]*Dipartimento di Scienze Matematiche, Fisiche e Informatiche, Università di Parma, I-43124 Parma, Italy*
[123]*INFN, Sezione di Milano Bicocca, Gruppo Collegato di Parma, I-43124 Parma, Italy*
[124]*Dipartimento di Ingegneria, Università del Sannio, I-82100 Benevento, Italy*
[125]*Università di Trento, Dipartimento di Fisica, I-38123 Povo, Trento, Italy*
[126]*INFN, Trento Institute for Fundamental Physics and Applications, I-38123 Povo, Trento, Italy*
[127]*Università degli Studi di Sassari, I-07100 Sassari, Italy*
[128]*INFN, Laboratori Nazionali del Sud, I-95125 Catania, Italy*
[129]*University of Portsmouth, Portsmouth PO1 3FX, United Kingdom*
[130]*West Virginia University, Morgantown, West Virginia 26506, USA*
[131]*The Pennsylvania State University, University Park, Pennsylvania 16802, USA*
[132]*Colorado State University, Fort Collins, Colorado 80523, USA*
[133]*Institute for Nuclear Research (Atomki), Hungarian Academy of Sciences, Bem tér 18/c, H-4026 Debrecen, Hungary*
[134]*CNR-SPIN, c/o Università di Salerno, I-84084 Fisciano, Salerno, Italy*
[135]*Scuola di Ingegneria, Università della Basilicata, I-85100 Potenza, Italy*
[136]*National Astronomical Observatory of Japan, 2-21-1 Osawa, Mitaka, Tokyo 181-8588, Japan*
[137]*Observatori Astronòmic, Universitat de València, E-46980 Paterna, València, Spain*
[138]*INFN Sezione di Torino, I-10125 Torino, Italy*
[139]*Indian Institute of Technology Bombay, Powai, Mumbai 400 076, India*
[140]*University of Szeged, Dóm tér 9, Szeged 6720, Hungary*
[141]*Delta Institute for Theoretical Physics, Science Park 904, 1090 GL Amsterdam, The Netherlands*
[142]*Lorentz Institute, Leiden University, P.O. Box 9506, Leiden 2300 RA, The Netherlands*
[143]*GRAPPA, Anton Pannekoek Institute for Astronomy and Institute for High-Energy Physics, University of Amsterdam, Science Park 904, 1098 XH Amsterdam, The Netherlands*

[144]Tata Institute of Fundamental Research, Mumbai 400005, India
[145]INAF, Osservatorio Astronomico di Capodimonte, I-80131 Napoli, Italy
[146]University of Michigan, Ann Arbor, Michigan 48109, USA
[147]American University, Washington, D.C. 20016, USA
[148]University of California, Berkeley, California 94720, USA
[149]Maastricht University, P.O. Box 616, 6200 Maryland Maastricht, The Netherlands
[150]Directorate of Construction, Services & Estate Management, Mumbai 400094 India
[151]University of Białystok, 15-424 Białystok, Poland
[152]King's College London, University of London, London WC2R 2LS, United Kingdom
[153]University of Southampton, Southampton SO17 1BJ, United Kingdom
[154]University of Washington Bothell, Bothell, Washington 98011, USA
[155]Institute of Applied Physics, Nizhny Novgorod, 603950, Russia
[156]Ewha Womans University, Seoul 03760, South Korea
[157]Inje University Gimhae, South Gyeongsang 50834, South Korea
[158]National Institute for Mathematical Sciences, Daejeon 34047, South Korea
[159]Ulsan National Institute of Science and Technology, Ulsan 44919, South Korea
[160]Bard College, 30 Campus Road, Annandale-On-Hudson, New York 12504, USA
[161]NCBJ, 05-400 Świerk-Otwock, Poland
[162]Institute of Mathematics, Polish Academy of Sciences, 00656 Warsaw, Poland
[163]Cornell University, Ithaca, New York 14850, USA
[164]Université de Montréal/Polytechnique, Montreal, Quebec H3T 1J4, Canada
[165]Lagrange, Université Côte d'Azur, Observatoire Côte d'Azur, CNRS, CS 34229, F-06304 Nice Cedex 4, France
[166]Hillsdale College, Hillsdale, Michigan 49242, USA
[167]Korea Astronomy and Space Science Institute, Daejeon 34055, South Korea
[168]Institute for High-Energy Physics, University of Amsterdam, Science Park 904, 1098 XH Amsterdam, The Netherlands
[169]NASA Marshall Space Flight Center, Huntsville, Alabama 35811, USA
[170]University of Washington, Seattle, Washington 98195, USA
[171]Dipartimento di Matematica e Fisica, Università degli Studi Roma Tre, I-00146 Roma, Italy
[172]INFN, Sezione di Roma Tre, I-00146 Roma, Italy
[173]ESPCI, CNRS, F-75005 Paris, France
[174]Center for Phononics and Thermal Energy Science, School of Physics Science and Engineering, Tongji University, 200092 Shanghai, People's Republic of China
[175]Southern University and A&M College, Baton Rouge, Louisiana 70813, USA
[176]Department of Physics, University of Texas, Austin, Texas 78712, USA
[177]Dipartimento di Fisica, Università di Trieste, I-34127 Trieste, Italy
[178]Centre Scientifique de Monaco, 8 quai Antoine Ier, MC-98000, Monaco
[179]Indian Institute of Technology Madras, Chennai 600036, India
[180]Université de Strasbourg, CNRS, IPHC UMR 7178, F-67000 Strasbourg, France
[181]Institut des Hautes Etudes Scientifiques, F-91440 Bures-sur-Yvette, France
[182]IISER-Kolkata, Mohanpur, West Bengal 741252, India
[183]Department of Astrophysics/IMAPP, Radboud University Nijmegen, P.O. Box 9010, 6500 GL Nijmegen, The Netherlands
[184]Kenyon College, Gambier, Ohio 43022, USA
[185]Whitman College, 345 Boyer Avenue, Walla Walla, Washington 99362 USA
[186]Hobart and William Smith Colleges, Geneva, New York 14456, USA
[187]Department of Physics, Lancaster University, Lancaster, LA1 4YB, United Kingdom
[188]OzGrav, Swinburne University of Technology, Hawthorn VIC 3122, Australia
[189]Trinity University, San Antonio, Texas 78212, USA
[190]Dipartimento di Fisica, Università degli Studi di Torino, I-10125 Torino, Italy
[191]Indian Institute of Technology, Gandhinagar, Ahmedabad, Gujarat 382424, India
[192]INAF, Osservatorio Astronomico di Brera sede di Merate, I-23807 Merate, Lecco, Italy
[193]Centro de Astrofísica e Gravitação (CENTRA), Departamento de Física, Instituto Superior Técnico, Universidade de Lisboa, 1049-001 Lisboa, Portugal
[194]Marquette University, 11420 West Clybourn Street, Milwaukee, Wisconsin 53233, USA
[195]Indian Institute of Technology Hyderabad, Sangareddy, Khandi, Telangana 502285, India
[196]INAF, Osservatorio di Astrofisica e Scienza dello Spazio, I-40129 Bologna, Italy
[197]International Institute of Physics, Universidade Federal do Rio Grande do Norte, Natal RN 59078-970, Brazil

[198]*Villanova University, 800 Lancaster Avenue, Villanova, Pennsylvania 19085, USA*
[199]*Andrews University, Berrien Springs, Michigan 49104, USA*
[200]*Carleton College, Northfield, Minnesota 55057, USA*
[201]*Department of Physics, Utrecht University, 3584CC Utrecht, The Netherlands*
[202]*Concordia University Wisconsin, 2800 North Lake Shore Drive, Mequon, Wisconsin 53097, USA*